\documentclass[12pt,a4paper]{iopart}
\pdfoutput=1
\usepackage{amssymb,amsfonts,graphicx}
\usepackage{color}

\begin{document}

\title[Statistical mechanics of quasi-geostrophic flows on a sphere]{Statistical mechanics of quasi-geostrophic flows on a rotating sphere.}
\author{C.Herbert$^{1,2}$, B.Dubrulle$^1$, P.H.Chavanis$^3$ and D.Paillard$^2$}
\address{$^1$ Service de Physique de l'Etat Condens\'e, DSM, CEA Saclay, CNRS URA 2464, Gif-sur-Yvette, France}
\address{$^2$ Laboratoire des Sciences du Climat et de l'Environnement, IPSL, CEA-CNRS-UVSQ, UMR 8212, Gif-sur-Yvette, France}
\address{$^3$ Laboratoire de Physique Th\'eorique (IRSAMC), CNRS and UPS, Universit\'e de Toulouse, 31062 Toulouse, France}
\ead{corentin.herbert@lsce.ipsl.fr}

\maketitle

\begin{abstract}

Statistical mechanics provides an elegant explanation to the
appearance of coherent structures in two-dimensional inviscid
turbulence: while the fine-grained vorticity field, described by the
Euler equation, becomes more and more filamented through time, its
dynamical evolution is constrained by some global conservation
laws (energy, Casimir invariants). As a consequence,
the coarse-grained vorticity field can be predicted through standard
statistical mechanics arguments (relying on the Hamiltonian structure
of the two-dimensional Euler flow), for any given set of the integral
constraints.

It has been suggested that the theory applies equally well to
geophysical turbulence; specifically in the case of the
quasi-geostrophic equations, with potential vorticity playing the role
of the advected quantity. In this study, we demonstrate analytically
that the Miller-Robert-Sommeria theory leads to non-trivial
statistical equilibria for quasi-geostrophic flows on
a rotating sphere, with or without bottom topography. We first consider
flows without bottom topography and with an infinite Rossby
deformation radius, with and without conservation of angular
momentum.  When the conservation of angular momentum is taken into account,
we report a case of second order phase transition associated with
spontaneous symmetry breaking. In a second step, we treat the general
case of a flow with an arbitrary bottom topography and a finite Rossby
deformation radius. Previous studies were restricted to flows in a planar domain 
with fixed or periodic boundary conditions with a beta-effect.

In these different cases, we are able to classify the statistical equilibria 
for the large-scale flow through their sole macroscopic features. 
We build the phase diagrams of the system and discuss the relations of the various statistical ensembles.

\end{abstract}


\submitto{Journal of Statistical Mechanics: Theory and Experiments.}

\tableofcontents

\clearpage

\section{Introduction}

An important characteristic of two-dimensional turbulent fluid flows
is the emergence of coherent structures: in the 80s, numerical
simulations \cite{McWilliams1984,Santangelo1989} showed that a
turbulent flow tends to organize itself spontaneously into large-scale
coherent vortices for a wide range of initial conditions and
parameters. Laboratory experiments reported similar observations
\cite{Couder1986,NguyenDuc1988,vanHeijst1989a,vanHeijst1989b}. Large-scale
coherent structures are also ubiquitous in planetary atmospheres and
in oceanography. Due to the long-lived nature of these structures, it
is very appealing to try to understand the reasons for their
appearance and maintenance through a statistical theory.

This endeavour is supported by theoretical arguments: as first noticed
by Kirchhoff, the equations for a perfect fluid flow can be recast in
a Hamiltonian form, which makes them \emph{a priori} suitable for
standard statistical mechanics treatments, as a Liouville theorem
automatically holds. The first attempt along these lines was Onsager's
statistical theory of point vortices \cite{Onsager1949}. One peculiar
outcome of Onsager's theory is the appearance of negative temperature
states at which large-scale vortices
form\footnote{Technically, the existence of negative temperatures
results from the fact that the coordinates $x$ and $y$ of the point
vortices are canonically conjugate. This implies that the phase space
coincides with the configuration space, so it is finite. This
leads to negative temperature states at high energies.}. The point
vortex theory was further developed by many authors
\cite{Montgomery1974,Kida1975,Pointin1976,Lundgren1977b,Frohlich1982,Benfatto1987}
and its relations with plasma physics
\cite{Joyce1973,Montgomery1974,Smith1990} and astrophysics \cite{Chavanis2002LNP,Eyink2006} was pointed out. The main problem with Onsager's
theory is that it describes a finite collection of point vortices and
not a continuous vorticity field. In particular, some invariant
quantities of perfect fluid flow are singular in the point vortex
description, and it is not easy to construct a continuum theory as a
limit of the point-vortex theory.

Subsequent attempts essentially considered truncations of the
equations of motion in the spectral space. Lee \cite{Lee1952} obtained
a Liouville theorem in the spectral phase space and constructed a
statistical theory taking into account only the conservation of
energy. Kraichnan built a theory on the basis of the conservation of
the quadratic invariants: energy and enstrophy
\cite{Kraichnan1967,Kraichnan1975}. The theory mainly predicts an
equilibrium energy spectrum corresponding to an equipartition
distribution. This spectrum has been extensively confronted with
experiments and numerical simulations
(e.g. \cite{Deem1971,Basdevant1975}) but the discussion remains open
\cite{ChorinBook}.

More recently, Miller \cite{Miller1990,Miller1992} and Robert \&
Sommeria \cite{Robert1991a,Robert1991b} independently developed a
theory for the continuous vorticity fields, taking into account all
the invariants of motion. Due to the infinite number of these
invariants, the rigorous mathematical justification is more elaborate
than previous approaches and relies on convergence theorems for Young
measures \cite{Michel1994b,Robert2000}. Miller \cite{Miller1992}
provides two alternative derivations, perhaps more heuristic, the
first one being based on phase space-counting ideas similar to
Boltzmann's classical equilibrium statistical mechanics (see also
Lynden-Bell \cite{LyndenBell1967}), while the second one uses a
Kac-Hubbard-Stratonovich transformation.  The MRS theory was checked
against laboratory experiments \cite{Monchaux2006} and numerical
studies \cite{Thess1994,Juttner1995} in a wide variety of publications
\cite{Tabeling2002}.

One of the main interest of applying statistical mechanical theories
to inviscid fluid flows is that it provides a very powerful tool to
investigate directly the structure of the final state of the flow,
regardless of the temporal evolution that leads to this final
state. From a practical point of view, such a tool would of course be
of great value as it is well-known that turbulence simulations are
very greedy in terms of computational resources. In some rare cases,
computations can be carried out analytically and it is even possible
to elucidate the final organization of the flow directly from the mean
field equations obtained from statistical mechanics. In any case, the
interest is also theoretical since equilibrium statistical mechanics
of inviscid fluid flows can be seen as a specific example of
long-range interacting systems \cite{DauxoisLRIbook}, whose
statistical mechanics is known to yield peculiar behaviors, in
phase transitions and ensemble inequivalence
\cite{Ellis2000,Ellis2002,Bouchet2005,Venaille2009,Venaille2011,Chavanis2006}. As an example, statistical mechanics provided valuable insight in the
understanding of a von Karman experiment, in particular in transitions
between different flow regimes \cite{Ravelet2004,Naso2010c},
fluctuation-dissipation relations \cite{Monchaux2008} and
Beltramization
\cite{Leprovost2005,Leprovost2006,Monchaux2006,Naso2010b}.

One particular area where avoiding long numerical turbulence
simulations would be highly beneficial is geophysics. Jupiter's great
red spot provides a prototypical example of application of statistical
mechanics to geophysical fluid dynamics
\cite{Miller1992,Michel1994a,Turkington2001,Bouchet2002,Chavanis2005},
in which valuable insight is gained from the statistical theory.  Even
before, the Kraichnan energy-enstrophy theory was extensively used to
discuss energy and enstrophy spectra in the atmosphere
\cite{Salmon1976,Frederiksen1980} and topographic turbulence
\cite{Herring1977,Merryfield1996}. However, only one study
\cite{Verkley2009b} considers the global equilibrium flow in a
spherical geometry, with encouraging results, but this study does not
investigate the structure of the flow in a systematic way. Statistical
mechanics of the continuous vorticity field conserving all the
invariants has also been applied to the Earth's oceans, focusing
either on small-scale parameterizations
\cite{Kazantsev1998,Polyakov2001,DiBattista1999,DiBattista2000,VenailleThesis}
or on meso-scale structures \cite{VenailleArxiv} (and in particular
the Fofonoff flow \cite{Venaille2009,Venaille2011,Naso2011}).

In this study, we investigate analytically the statistical equilibria
of the large-scale general circulation of the Earth's atmosphere,
modelled by the quasi-geostrophic equations, taking into account the
spherical geometry (with possible bottom topography) and the full
effect of rotation, in the framework of the MRS theory.  More
precisely, we show that in the absence of a bottom topography, due to
the spherical geometry, the solution to the statistical mechanics
problem can be derived in a very simple way. The result is, however,
highly non trivial because, when the conservation of angular momentum
is properly accounted for, it leads to a second order phase transition
associated with a spontaneous symmetry breaking. Since all the
previous studies used a $\beta$-effect instead of the full Coriolis
parameter and focused on rectangular bounded regions rather than on
the full sphere, this simple solution was not noticed before. We draw
the phase diagrams of the system in both microcanonical and
grand-canonical ensembles. The relations between the two statistical
ensembles is described in detail and we present a refined notion of
marginal equivalence of ensemble (see also \cite{Herbert2011c}).  In
the presence of a bottom topography, we obtain semi-persistent
equilibria reminiscent of the structures observed in the
atmosphere. They correspond to saddle points of entropy. Strictly
speaking, they are unstable since they can be destabilized by certain
infinitesimal perturbations belonging to particular subspaces of the
dynamical space: for these saddle points of the entropy surface, there
is at least one direction along which the entropy increases while the
constraints remain satisfied. However, it may take a long time before
the system spontaneously generates these perturbations. Therefore,
these states may persist for a long time before finally being
destabilized \cite{Naso2010a}. In the atmospheric context, these semi-persistent
equilibrium states could account for situations of atmospheric
blocking where a large scale structure can form for a few days before
finally disappearing.

In section \ref{genstatmachsection}, we present the general
statistical mechanics of the quasi-geostrophic equations. In section
\ref{notopoRinftysection} we obtain the structure of the equilibrium
mean flow in the particular case of a sphere without bottom topography
in the limit of infinite Rossby deformation radius,
with and without conservation of angular momentum. In
section \ref{generalQG1section}, we examine the effect of the bottom
topography and of the Rossby deformation radius. Section
\ref{discussionsection} presents a discussion of the obtained results
and a comparison with previously published results, while conclusions are presented
in section \ref{conclusionsection}.

\section{Statistical Mechanics of the quasi-geostrophic equations}\label{genstatmachsection}

\subsection{Definitions and notations}

We consider here an incompressible, inviscid, fluid on the two
dimensional sphere $S^2$ (denoted $D$ to keep notations simple). The
coordinates are $(\theta, \phi)$ where $\theta \in [0,\pi]$ is the
polar angle (the latitude is thus $\pi/2-\theta$) and $\phi \in
[0,2\pi]$ the azimuthal angle.  In the following, for any quantity
$A$, we note $\langle A \rangle$ its average value over the whole
domain:
\begin{equation}
\langle A \rangle = \frac{\int_D A({\bf r}) d^2{\bf r}}{\int_D d^2{\bf r}}.
\end{equation}
We introduce the eigenvectors of the Laplacian $\Delta$ on the sphere. These are 
the spherical harmonics $Y_{nm}$ with eigenvalues $\beta_n$:
\begin{eqnarray}
Y_{nm}(\theta,\phi)&=&\sqrt{\frac{2n+1}{4\pi} \frac{(n-m)!}{(n+m)!} } P_n^m(\cos \theta) e^{im\phi},\\
\Delta Y_{nm} &=& \beta_n Y_{nm},
\end{eqnarray}
where $P_n^m$ are the associated Legendre polynomials and $\beta_n=-n(n+1)$ \cite{GradshteynRyzhik}. The scalar product on the vector space of complex-valued functions on the sphere $S^2$ is defined as usual as
\begin{equation}
\langle f | g \rangle = \int_0^{2\pi} d\phi \int_0^\pi d\theta \sin \theta \overline{f(\theta,\phi)} g(\theta,\phi),
\end{equation}
where the bar denotes complex conjugaison, so that the spherical harmonics form an orthonormal basis of the Hilbert space $L^2(S^2)$:
\begin{equation}
\langle Y_{nm} | Y_{pq} \rangle = \delta_{np} \delta_{mq}.
\end{equation}
Note that $\langle f | g \rangle = 4\pi \langle \bar{f} g \rangle$.

For applications to the Earth, we shall take the inverse of the Earth's rotation rate $\Omega$ as the time unit and we set $\tilde{r}=r/R_T$ in the radial direction so that the Earth mean radius $R_T$ is the length unit. Hence all the analytical calculations are carried out on the unit sphere $S^2$, while we retain the $\Omega$ dependance in the calculations to stress the effect of rotation in the formulae, even though for numerical applications, we will always take $\Omega=1$.

\subsection{The quasi-geostrophic equations}\label{qgeqssection}

We consider here the simplest model for geophysical flows: the
one-layer quasi-geostrophic equations, also called the equivalent
barotropic vorticity equations. We assume that the velocity field $\bf
v$ satisfies the incompressibility condition $\nabla . {\bf v}=0$, so
that we can introduce a stream function $\psi$ such that ${\bf
v}=-{\bf \hat{r}} \times {\bf \nabla} \psi$, and define the potential
vorticity as
\begin{equation}
q=-\Delta \psi +h +\frac{\psi}{R^2},
\end{equation}
where $h$ is the topography and $R$ the Rossby deformation radius \cite{PedloskyGFD}. The evolution of the potential vorticity is given by the quasi-geostrophic equation
\begin{equation}\label{QGeq1}
\partial_t q + {\bf v}\cdot \nabla q=0.
\end{equation}
In other words, the flow conserves potential vorticity. Together with the fact that the flow is incompressible, this implies the conservation of the integral of any function of potential
vorticity $I_g = \int_D g(q) d^2{\bf r}$, called \emph{Casimir
invariants} ($g$ being an arbitrary function). In particular, any
moment $\Gamma_n = \int_D q^n d^2{\bf r}$ of the potential vorticity
is conserved. $\Gamma_1$ will be called here the \emph{circulation}
and $\Gamma_2$ the \emph{potential enstrophy}. The energy, given by
\begin{equation}
E = \frac{1}{2} \int_D (q-h)\psi d^2{\bf r} = \frac{1}{2}\int_D \left((\nabla \psi)^2 + \frac{\psi^2}{R^2} \right)d^2{\bf r},
\end{equation}
is also conserved. Finally, due to the spherical
symmetry, one may also consider a supplementary invariant: the
integral over the domain of the vertical component of angular
momentum
\begin{equation}
L=\int_D u \sin \theta d^2{\bf r}=\int_D q \cos \theta d^2 {\bf r},
\end{equation}
where $u=-\partial_\theta \psi$ is the zonal component of velocity. In \ref{solidbodyrotationsappendix}, we show that, for a solid-body rotation, the dynamical invariants $E$ and $L$
are not independent: they obey a relation of the form $E=E^*(L)$, with
$E^*(L)=3L^2/4$. We also show (\ref{ELappendix}) that, for any flow,
$E\geq E^*(L)$.

For fluid motion on a rotating sphere, the term $h$ includes the
Coriolis parameter $f=2\Omega \cos \theta$. Following
\cite{Verkley2009b}, the general form we will consider here is
$h=f+f\frac{h_B}{H_A}$ with $h_B$ the bottom topography and $H_A$ the
average height of the fluid. The relative vorticity is $\omega=-\Delta
\psi$ and the absolute vorticity $\omega+f$. In the limit of infinite
Rossby deformation radius ($R=\infty$) and no topography ($h=0$), we
recover the 2D Euler equations. Introducing the Poisson brackets on the
sphere 
\begin{equation}
\{ A , B \} =
\frac{1}{r^2 \sin \theta} \left( \frac{\partial A}{\partial
\phi}\frac{\partial B}{\partial \theta} - \frac{\partial A}{\partial
\theta}\frac{\partial B}{\partial \phi} \right),
\end{equation}
the quasi-geostrophic equation (\ref{QGeq1}) reads
\begin{equation}
\partial_t q + \{ q, \psi \} =0. \label{qpsipoissoneq}
\end{equation}
It is well known in the case of the Euler (or quasi-geostrophic in a
planar domain) equations that the Poisson bracket form implies that
the steady states of the equations correspond to $q=F(\psi)$ with $F$
an arbitrary function. In fact, due to the particular geometry
considered here, the form of the steady-states must be slightly
refined. Let us consider solutions of the quasi-geostrophic equations
of the form $q(\theta,\phi,t)=q(\theta,\phi-\Omega_L
t)$. Substituting this relation into equation (\ref{qpsipoissoneq}), we obtain
\begin{eqnarray}
-\Omega_L \frac{\partial q}{\partial \phi} + \{ q, \psi \}=\{ q , \Omega_L \cos \theta \} + \{ q, \psi \} =0,
\end{eqnarray}
so that $q=F(\psi+\Omega_L \cos \theta )$, with $F$ an arbitrary
function. This is the general form of the solutions of the
quasi-geostrophic equations which are stationary in a frame rotating
with angular velocity $\Omega_L$ with respect to the initial reference
frame (which rotates with angular velocity $\Omega$). When
$\Omega_L=0$, we recover the previous $q-\psi$ relationship. However,
due to the spherical symmetry, there is no reason to select the
reference frame $\Omega_L=0$ \emph{a priori}.

In the next section, we show that statistical mechanics allows to select a particular function $F$ on the grounds that it is the most probable equilibrium state respecting the constraints.

\subsection{Maximum entropy states}

If we were to inject a droplet of dye in a turbulent two-dimensional
flow, we would observe a complex mixing where the originally regular
patch of dye turns into finer and finer filaments as time goes
on. After a while, the filaments are so intertwined that the dye seems
homogeneously distributed over the fluid to a human eye: the
coarse-grained dye concentration is homogeneous. In the
quasi-geostrophic equations, it is potential vorticity that is mixed
by the flow (equation (\ref{QGeq1})). The crucial difference is that
the advected quantity is no longer a passive tracer but plays an
active role in the dynamics. Due to the conservation constraints
associated to the quasi-geostrophic equations, the potential vorticity
mixing will not lead to an homogeneous coarse-grained distribution. In
particular, the energy constaint prevents complete mixing. We wish to
determine what this final coarse-grained state will be, regardless of
the details of the fine-grained structure of the potential vorticity
field.  Analogously to classical statistical mechanics
\cite{Jaynes1957a,BalianBook,Balian2005}, after identifying the
correct description for microstates (exact fine-grained vorticity
field) and macrostates (the coarse-grained vorticity field,
mathematically represented as a Young measure), one selects the
macrostate that maximizes the statistical entropy subject to the
relevant macroscopic constraints (conserved quantities), as developed
by Miller and Robert \cite{Miller1990, Miller1992, Robert1991a, Robert1991b}. The
underlying fondamental property is that an overwhelming majority of
microstates lie in the vicinity of the equilibrium macrostate. The
implicit separation of scales between microstates and macrostates
implies that the contributions of the small-scale fluctuations of
vorticity are discarded in the macroscopic quantities (strictly speaking, 
this is true for the energy but not for the Casimirs: computing 
the moments of the vorticity distribution using the fine-grained 
distribution or the coarse-grained distribution yields different 
results). As a consequence, the
Miller-Robert-Sommeria (MRS) theory is a mean-field theory
\cite{Jung2006, Miller1992}. Note also that albeit all the dynamically
conserved quantities of the equations are imposed as constraints in
the statistical mechanics, the topological constraints are not
conserved: a connected vorticity domain should remain connected
through time, while in the statistical mechanics the only thing that
is conserved is the area of this domain.

At the microscopic level, the potential vorticity is fully determined
by the initial conditions and the evolution equation (\ref{QGeq1}). At
the macroscopic level, we consider the coarse-grained potential
vorticity $q$ as a random variable with probability distribution
$\rho$: the probability that the potential vorticity has the value
$\sigma$ with an error $d\sigma$ at point $\bf r$ is $\rho({\bf r},
\sigma)d\sigma$. The potential vorticity distribution $\rho({\bf
r},\sigma)$ characterizes the macroscopic state. The potential
vorticity distribution must satisfy the normalization condition $\int
\rho({\bf r},\sigma) d\sigma =1$ at each point of the domain, and the
mean value of the potential vorticity is given by $\overline{q}=\int
\rho({\bf r},\sigma)\sigma d\sigma$. We introduce a stream function
$\psi$ corresponding to the ensemble-mean potential vorticity through
$\overline{q} = - \Delta \psi + \frac{\psi}{R^2} + h$. The statistical
entropy of the probability distribution $\rho$ is 
$$S[\rho] = - \mbox{
Tr} \left(\rho \ln \rho \right)= -
\int_{-\infty}^{+\infty} d\sigma \int_D d^2{\bf r} \rho({\bf
r},\sigma) \ln \rho({\bf r},\sigma).$$
We are looking for the probability distribution $\rho$ that maximizes
the statistical entropy functional $S[\rho]$ subject to the
constraints mentioned in section \ref{qgeqssection}: global
conservation of energy and Casimir functionals. The conservation of
all the Casimirs is equivalent to the conservation of the area of each
potential vorticity level $\gamma(\sigma) = \int_D \rho({\bf
r},\sigma) d^2{\bf r}$. Hence the statistical equilibria must satisfy
\begin{eqnarray}
\fl
\delta S - \tilde{\beta} \delta E - \int \tilde{\alpha}(\sigma)\delta \gamma(\sigma)d\sigma 
-\tilde{\mu} \int_D \delta \left(\int \sigma \rho({\bf r},\sigma) d \sigma\right) \cos \theta d^2{\bf r} 
\nonumber\\
-\int_D \zeta({\bf r}) \delta \left( \int \rho({\bf r},\sigma) d\sigma\right) d^2{\bf r} = 0,
\end{eqnarray}
where $\tilde{\beta},\tilde{\alpha}(\sigma),\tilde{\mu}$ and $\zeta({\bf r})$ are respectively the Lagrange multipliers associated with the conservation of energy, potential vorticity levels, angular momentum, and normalization.

The resulting potential vorticity probability density is the Gibbs
state
\begin{equation}
\rho({\bf r},\sigma) = \frac{1}{Z} g(\sigma) e^{-\tilde{\beta} \sigma \psi-\tilde{\mu}\sigma\cos\theta},
\end{equation}
where $g(\sigma)=e^{-\tilde{\alpha}(\sigma)}$ and $Z=e^{1+\zeta({\bf
r})}$. Due to the normalization condition, the partition function $Z$
is also given by
\begin{equation}
Z=\int g(\sigma) e^{-\tilde{\beta}\sigma \psi -\tilde{\mu}\sigma \cos\theta} d\sigma,
\end{equation}
and the ensemble-mean potential vorticity satisfies the usual relation
\begin{equation}
\overline{q}=-\frac{1}{\tilde{\beta}} \frac{\partial \ln Z}{\partial \psi}.
\end{equation}
The right-hand side of this equation is a certain function $F$ of the relative stream function
$\psi_*=\psi + \frac{\tilde{\mu}}{\tilde{\beta}}\cos\theta$. Hence, for given values
of the Lagrange multipliers $\tilde{\alpha}$ and $\tilde{\beta}$, the
statistical entropy maximization procedure selects a functional
relationship between potential vorticity and relative stream function at
steady-state: $\overline{q} = F_{\tilde{\beta},
\tilde{\alpha},\tilde{\mu}}\left(\psi +
\frac{\tilde{\mu}}{\tilde{\beta}}\cos\theta\right)$. This describes a flow rotating with angular velocity $\Omega_L=\tilde\mu/\tilde\beta$ with respect to the terrestrial frame. Therefore, statistical mechanics selects steady states of the QG equations in a rotating frame. The resulting mean field equation
is simply
\begin{equation}\label{generalstateeq}
-\Delta \psi + \frac{\psi}{R^2}+h = F\left(\psi+\frac{\tilde{\mu}}{\tilde{\beta}}\cos\theta\right).
\end{equation}
This is the general mean field equation for equilibrium states of the
quasi-geostrophic equations. In the limit $R\to \infty,
h=0,\tilde{\mu}=0$, one recovers the well-known mean field equation
for the Euler equation.  Note that in the case of only two potential
vorticity levels $\sigma_1$ and $\sigma_{-1}$,
the partition function $Z$ is simply $Z=g(\sigma_1)e^{-\tilde{\beta}
\sigma_1 \psi_*}+g(\sigma_{-1})e^{-\tilde{\beta} \sigma_{-1} \psi_*}$, so
that after straightforward computations, we find the $q-\psi$
relationship $\bar{q}=B-A\tanh\left(\alpha+A \tilde{\beta}\psi_*\right)$
with $e^{-2\alpha}=\frac{g(\sigma_1)}{g(\sigma_{-1})}$, $B =
\frac{\sigma_1 + \sigma_{-1}}{2}$, $A = \frac{\sigma_1 -
\sigma_{-1}}{2}$.  With $A=1$ and
$\tilde{\beta}=-\frac{C}{R^2}$, we recover the $q-\psi$ relationship
\cite{Bouchet2002}:
\begin{equation}\label{stateeqBouchet}
\bar{q} = B - \tanh \left( \alpha - \frac{C \psi_*}{R^2}\right).
\end{equation}

To determine the statistical equilibrium state, we have to solve the
mean field equation (\ref{generalstateeq}), relate the Lagrange
multipliers to the constraints and study the stability of the
solutions (whether they are entropy maxima or saddle points). If
several entropy maxima are found for the same parameters (conserved
quantities), we must distinguish metastable states (local entropy maxima)
from fully stable states (global entropy maxima).

\subsection{The linear $q-\psi$ relationship}

In practice, the mean field equation (\ref{generalstateeq}) is
difficult to solve because the function $F$ is in general nonlinear
due to the conservation of all the moments of the fine-grained
potential vorticity. Besides, it is generally difficult to relate the
Lagrange multipliers to the conserved quantitites. The two levels system
of \cite{Bouchet2002} provides an example of a case where it is
possible to write down explicitly the $q-\psi$ relationship but the
meanfield equation (\ref{stateeqBouchet}) is not analytically solvable
without further approximations. Nevertheless, efficient numerical
methods do exist, like for instance the algorithm of Turkington and
Whitaker
\cite{Turkington1996} or the method of relaxation equations
\cite{Robert1992,Chavanis2009}.

To go further with analytical methods, a common solution is to
linearize the $q-\psi$ relationship. Several justifications of this
procedure can be given, which can be grossly classified in two types
of approaches. In the first approach, one simply discards the effect of the
high-order fine-grained potential vorticity moments (it would be
possible to include them one by one hierarchically), while in the
second approach, their effect is prescribed through a gaussian \emph{prior}
distribution for small-scale potential vorticity (in the general
theory, one can specify a non-gaussian prior, which would lead to a
nonlinear $q-\psi$ relationship).

\begin{itemize}
\item In the limit of strong mixing $\tilde{\beta} \sigma \psi \ll 1$, 
the argument of the exponential in the partition function is small 
and a power expansion of $Z$ can be carried out. The rigorous 
computation is presented in \cite{Chavanis1996} and yields a 
linear $\overline{q}-\psi$ relationship. This power series expansion 
can be done at virtually any order. At first-order, equation (\ref{generalstateeq}) becomes identical
to the mean field equation obtained by minimizing the coarse-grained
enstrophy $\Gamma_2^{cg} = \int_D \bar{q}^2\, d^2{\bf r}$ with fixed
energy, circulation and angular momentum. The value of the
fine-grained enstrophy $\Gamma_2^{fg} =
\int_D \bar{q^2}\, d^2{\bf r}$ is fixed by the initial condition and we always
have $\Gamma_2^{cg} \leq \Gamma_2^{fg}$.  The strong  mixing limit thus
corresponds to cases where the energy, circulation, angular momentum (called
\emph{robust invariants} because they are expressed in terms of the
coarse-grained potential vorticity) and fine-grained enstrophy are the
only important invariants and the higher-order moments of the
fine-grained enstrophy (called \emph{fragile invariants})
do not play any role. This can be seen as a form of justification in the framework of statistical mechanics of inviscid fluids of the early phenomenological minimum enstrophy principle suggested by \cite{Bretherton1976}, \cite{Matthaeus1980} and \cite{Leith1984} on the basis of the inverse cascade of Batchelor \cite{Batchelor1969} for finite viscosities. The connection between the inviscid statistical theory and the phenomenological \emph{selective decay} approach is discussed at length in \cite{Chavanis1996}, \cite{Chavanis1998b} and \cite{Brands1999}.

\item For any given energy, one can find a vorticity level distribution $\gamma(\sigma)$ such that the function $F$ is linear. This corresponds to a gaussian $g(\sigma)$ (see \cite{Miller1992}). Indeed, if $g(\sigma)=\frac{1}{\sqrt{2\pi \eta}} e^{-\frac{(\sigma - \sigma_m)^2}{2\eta}}$ is a Gaussian with mean value $\sigma_m$ and standard deviation $\eta$, the analytical computation of the partition function is straightforward
\begin{equation}
Z=e^{\frac{\eta}{2}\tilde{\beta}^2\psi_*^2 -
\sigma_m\tilde{\beta} \psi_*-\frac{\sigma_m^2}{2\eta}},
\end{equation}
and the mean flow satisfies the equation
\begin{equation}\label{stateeqationEHTgaussian}
\overline{q} = - \frac{1}{\tilde{\beta}} \frac{\partial \ln Z}{\partial \psi} = - \eta \tilde{\beta}  \psi_*  + \sigma_m.
\end{equation}
Furthermore, if the flow maximizes $S[\bar{q}] = - \frac{1}{2}\int_D \bar{q}^2\, d^2{\bf r}$ 
at fixed energy and circulation, then it is granted to be thermodynamically stable 
in the MRS sense \cite{Bouchet2008,Venaille2009} (see also \cite{Chavanis2009}).
In the approach of Ellis, Haven and Turkington \cite{Ellis2002}, 
$g(\sigma)$ is interpreted as a  \emph{prior distribution} for the high-order moments of
potential vorticity (fragile constraints): arguing that real flows are
subjected to forcing and dissipation at small scales, Ellis {\it et al.} \cite{Ellis2002} 
objected that conservation of the fragile
constraints (which depend on the fine-grained field) is probably
irrelevant. They suggested to treat these constraints canonically by
fixing the Lagrange multiplier $\alpha_n$ instead of $\Gamma_n^{fg}$
itself. Chavanis \cite{Chavanis2008b,Chavanis2010c} showed that this is
equivalent to maximizing a relative entropy $S_\chi = -\int \rho \ln
\frac{\rho}{\chi}\, d^2{\bf r}d\sigma$ with a prescribed \emph{prior distribution} $\chi(\sigma)$
 for the small-scale potential vorticity. The ensemble-mean
coarse-grained potential vorticity then is a maximum of a generalized
entropy functional $S[\bar{q}] = - \int_D C(\bar{q})\, d^2{\bf r}$
with fixed values of the robust invariants (energy, circulation and angular momentum),
where $C$ is a convex function determined by the prior $\chi$
\cite{Chavanis2008b}. The linear ${q}-\psi$ relation
(\ref{stateeqationEHTgaussian}) corresponds to a gaussian prior $\chi$
and, in this case, the generalized entropy is minus the coarse-grained
enstrophy, i.e. $S[\bar{q}] = -\frac{1}{2}\int_D \bar{q}^2\, d^2{\bf r}$.

\end{itemize}

As an intermediate case, Naso \emph{et al.} \cite{Naso2010a} take 
up the argument that the conservation of some Casimirs is broken by
small-scale forcing and dissipation, but instead of prescribing a
prior small-scale vorticity distribution, they suggest that the relevant
invariants to keep are determined directly by forcing and
dissipation (which, on average, equilibrate so that the system reaches
a quasi-stationary state). They show that maximizing the
Miller-Robert-Sommeria entropy with fixed energy, circulation, angular momentum and
fine-grained enstrophy is equivalent (for what concerns the
macroscopic flow) to minimizing the coarse-grained enstrophy at fixed
energy, circulation and angular momentum. Furthermore, the fluctuations around this
macroscopic flow are gaussian.

Note that one does not necessarily need to justify physically the 
linear $q-\psi$ relationship: we may argue that we are just studying a
subset of the huge and notoriously difficult to compute class of MRS
statistical equilibria.

In the following sections, we shall study the mean equilibrium flow
for the quasi-geostrophic equations on the sphere based on these
equivalent formulations of the variational problem: we use the
generalized entropy $S[\bar{q}]=-\frac{1}{2}\langle \bar{q}^2 \rangle
= -\Gamma_2^{cg}[\bar{q}]/(2|D|)$ where $|D|=4\pi$ is the area of the
unit sphere\footnote{In the following, the bar on $\bar{q}$ will be dropped for
convenience.}, and we consider maxima of this functional with fixed
energy $E$, circulation $\Gamma$ and angular momentum $L$. Hence, our study
is restricted to the case of a linear $q-\psi$ relationship. Note that
from the physical point of view, it is not an irrelevant restriction
as a large class of geophysical flows are described by linear $q-\psi$
relationships, like the Fofonoff flows in oceanography
\cite{Fofonoff1954}. Besides, a strong point is that, in this limit, as
we will see in the following sections, the analytical methods allow us
to study a large family of metastable states that may be relevant for
the atmosphere.

\subsection{Statistical ensembles and variational problems}

In this study, we shall consider the maximization of the generalized
entropy $S[q]=-\frac{1}{2}\langle q^2 \rangle$, as explained in the
previous section, with either fixed energy and circulation, or fixed
energy, circulation, and angular momentum. The corresponding
variational problems can be written as
\begin{equation}
{\cal S}(E,\Gamma) = \max_q \{ S[q]  | E[q]=E, \Gamma[q]=\Gamma \},
\label{vw1}
\end{equation}
and
\begin{equation}
{\cal S}(E,\Gamma,L) = \max_q \{ S[q]  | E[q]=E, \Gamma[q]=\Gamma, L[q]=L \}.
\label{vw2}
\end{equation}
These constrained variational problems correspond to the microcanonical
ensemble and the function ${\cal S}$ is called the entropy. We shall
also consider the dual variational problems with relaxed constraints
\begin{equation}
{\cal J}(\beta,\alpha) = \max_q \{ S[q]  -\beta E[q] - \alpha \Gamma[q] \},
\label{vw3}
\end{equation}
and
\begin{equation}
{\cal J}(\beta,\alpha,\mu) = \max_q \{ S[q]  - \beta E[q] - \alpha \Gamma[q] - \mu L[q]\}.
\label{vw4}
\end{equation}
In both cases, the corresponding statistical ensemble will be termed
\emph{grand-canonical}\footnote{In the literature, the variational
problem ${\cal J}(\beta,\alpha,\mu)$ is sometimes called
\emph{grand-grand-canonical} \cite{Chavanis2009}. Here, the distinction between
grand-canonical and grand-grand-canonical will always be clear because
the two ensembles are considered in separate sections. Hence, we shall
keep the vocabulary to a minimum and call both variational problems
\emph{grand-canonical}. }. The function ${\cal J}$ will be called
\emph{grand-potential} in both cases\footnote{The same remark as above
holds for \emph{grand-potential} and \emph{grand-grand-potential}}.

Clearly, the critical points of the microcanonical and grand-canonical
variational problems are the same, due to the Lagrange multiplier
theorem. However the nature of these critical points (maximum,
minimum, saddle point) can differ. Nevertheless, it is straightforward
to convince oneself that a maximum of the relaxed variational problem
is also a maximum of the constrained variational problem
(grand-canonical stability implies microcanonical stability) and that
a saddle point of the constrained variational problem is necessarily
a saddle point of the relaxed variational problem (microcanonical
instability implies grand-canonical instability). More detailed
relationships between the constrained and relaxed variational problems
can be found in \cite{Ellis2000,Ellis2002,Bouchet2008,Chavanis2009}.

It is thus possible that a maximum in the constrained variational
problem (that is an equilibrium state in the microcanonical ensemble)
may not be reached in the grand-canonical ensemble. Such a situation
happens when no grand-canonical equilibrium has the prescribed energy,
circulation, and angular momentum if relevant. In this case, we speak
of \emph{ensemble inequivalence}. More precisely, we have just
described ensemble inequivalence at the \emph{macrostate level}
\cite{Ellis2000,Touchette2004}. Another characterization of ensemble
inequivalence, referred to as ensemble inequivalence at the
\emph{thermodynamic level}, is linked to the concavity of the entropy
${\cal S}$ \cite{Ellis2000,Ellis2002,Touchette2004}. Indeed, the
functions ${\cal S}$ and ${\cal J}$ are linked by Legendre-Fenchel
transformations. Using convex analysis \cite{RockafellarBook}, it can
be proved that the transformation is invertible only when ${\cal S}$
is a concave function. A particular indicator of ensemble
inequivalence is the microcanonical specific heat. It is easily proved
that when computed with the grand-canonical probability distribution,
the specific heat is always positive. On the other hand, no such
result holds in the microcanonical ensemble. Thus, a negative
microcanonical specific heat indicates ensemble inequivalence
\cite{LyndenBell1968,Thirring1970,Barre2001,Chavanis2006}.

In practice, the microcanonical ensemble is the natural one to treat
problems where the system is large enough to be considered isolated,
as in astrophysics or geophysical flows. Yet, it is always more
convenient mathematically to deal with a relaxed variational problem
than directly with the constrained variational problem. For this
reason, it is customary to consider canonical or grand-canonical
ensembles even in these cases. Meanwhile, one must keep in mind that
the physical interpretation of the canonical or grand-canonical
ensemble may not be straightforward. In our case, it is not easy to
see what physical object would play the role of a reservoir of energy,
circulation or angular momentum. In other words, it is
not clear how the Lagrange multipliers $\beta, \alpha$ and $\mu$ are fixed, and which physical quantity they represent.
Strictly speaking, the canonical and grand-canonical ensembles 
are physically relevant only when they are equivalent with the 
microcanonical ensemble (or when there is a good physical reason 
to be otherwise). On the contrary, the microcanonical ensemble is always physically relevant.

{\it Remark}: We have justified the variational principles 
(\ref{vw1})-(\ref{vw4}) as conditions of thermodynamical stability. We note 
that the very same  variational principles can also be interpreted as
 sufficient conditions of nonlinear dynamical stability with respect to 
the QG equations \cite{Ellis2002,Chavanis2009}. Therefore, the stable states that 
we shall determine are both dynamically and thermodynamically stable.


\section{The equilibrium mean flow in the barotropic case ($R=\infty$) without bottom topography ($h_B=0$)}
\label{notopoRinftysection}

\subsection{Fixed energy and circulation}

In this section, we consider the following variational problems
\begin{eqnarray}\label{minensvarprob1}
{\cal S}(E,\Gamma) = \max_q \{ S[q]  | E[q]=E, \Gamma[q]=\Gamma \},
\end{eqnarray}
and
\begin{eqnarray}\label{minensvarprob2}
{\cal J}(\beta,\alpha) = \max_q \{ S[q]  -\beta E[q] - \alpha \Gamma[q] \},
\end{eqnarray}
where
\begin{equation}
S[q]=-\frac{1}{2}\langle q^2\rangle, \quad E[q]=\frac{1}{2}\langle(q-f)\psi\rangle, \quad \Gamma[q]=\langle q \rangle,
\end{equation}
are respectively the (normalized) entropy, energy and circulation. The critical points satisfy
\begin{equation}
\delta S-\beta \delta E - \alpha \delta \Gamma=0,
\end{equation}
where $\alpha$ and $\beta$ are Lagrange multipliers associated with the conservation of circulation and energy, respectively. This leads to the linear $q-\psi$ relationship
\begin{equation}
q=-\beta \psi - \alpha.
\end{equation}
Averaging over space, we obtain $\langle q \rangle= - \beta \langle \psi \rangle - \alpha$, 
but $\Gamma=\langle q \rangle = \langle -\Delta \psi + f\rangle=0$ since 
the integral of $\Delta \psi$ is the circulation of the velocity on 
the domain boundary, hence vanishing in the case of the sphere. Thus, $\alpha=-\beta\langle\psi\rangle$ and  the $q-\psi$ relationship reads
\begin{equation}
q=- \Delta \psi + f=-\beta \left(\psi -\langle \psi \rangle\right).
\end{equation}
If we let $\phi = \psi - \langle \psi \rangle$ such that $\langle \phi \rangle =0$, the mean field equation reduces to the simple Helmholtz equation
\begin{equation}\label{stateHelmholtzEq}
\Delta \phi - \beta \phi = f.
\end{equation}
Here, we note that $f=2\Omega \cos \theta= 2 \Omega \sqrt{\frac{4\pi}{3}}Y_{10}(\theta,\phi)$ is an eigenvector of the Laplacian on the sphere (with eigenvalue $\beta_1=-2$). Therefore if we introduce the operator $A_\beta = \Delta - \beta I$, $f$ satisfies $A_\beta f = - (\beta-\beta_1)f$. To solve equation (\ref{stateHelmholtzEq}), we have to distinguish two cases, depending whether the inverse temperature is an eigenvalue of the Laplacian or not \cite{Chavanis1996}.

Note that the Lagrange multiplier $\alpha$ is only related to the mean value 
of the stream function on the domain, $\langle \psi \rangle$, which has no effect 
on the structure of the flow (the stream function is defined up to an unimportant additive constant). 
This is not surprising as in the case considered in this section ($R=+\infty, h=f$), 
we must always have $\Gamma=0$. Therefore, neither $\alpha$ nor $\Gamma$ will intervene in the following discussion.
To keep the notations simple in the sequel, we shall make the gauge choice $\langle \psi \rangle =0$ 
and identify $\psi$ and $\phi$. This also implies $\alpha=0$.

\subsubsection{Case $\beta \notin \mbox{ Sp } \Delta$: the continuum solution}

\begin{figure}
\begin{centering}
\includegraphics[width=0.7\textwidth]{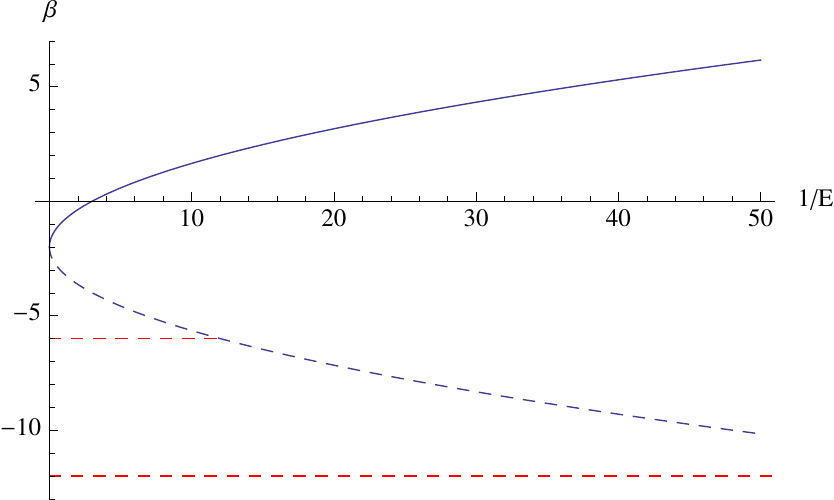}
\caption{Caloric curve $\beta(1/E)$ in the case where only the energy and circulation are conserved. The upper part (blue solid line) corresponds to real entropy maxima while the lower part (dashed lines) corresponds to saddle points. Horizontal lines indicate the position of the eigenvalues of the Laplacian, and therefore correspond to plateaux of (saddle) solutions.}\label{caloriccurveEGfig}
\end{centering}
\end{figure}

In this case, equation (\ref{stateHelmholtzEq}) reduces to $A_\beta \left(\psi +\frac{f}{\beta-\beta_1}\right)=0$, but $A_\beta$ is invertible, hence
\begin{equation}
\label{mo1}
\psi=-\frac{f}{\beta-\beta_1},
\end{equation}
and
\begin{equation}
q=\frac{\beta f}{\beta-\beta_1}.
\label{mo2}
\end{equation}
Using equations (\ref{mo1}) and (\ref{mo2}), and  $\langle f^2\rangle=\frac{4}{3}\Omega^2$, the 
equilibrium energy and entropy are easily computed. We get
\begin{equation}
E(\beta)=\frac{1}{2}\langle(q-f)\psi\rangle =-\beta_1\frac{\langle f^2 \rangle}{2(\beta-\beta_1)^2}
=\frac{4\Omega^2}{3(\beta-\beta_1)^2},
\end{equation}
and
\begin{equation}
S=-\frac{1}{2}\langle q^2 \rangle
=-\frac{\beta^2}{2(\beta-\beta_1)^2}\langle f^2 \rangle
=- \frac{2 \Omega^2 \beta^2}{3(\beta-\beta_1)^2}.
\end{equation}
The relation between the energy $E$ and the Lagrange multiplier $\beta$ can be solved:
\begin{equation}
\beta(E)=\beta_1\pm \sqrt{\frac{4\Omega^2}{3E}},
\end{equation}
(see the caloric curve on figure \ref{caloriccurveEGfig}),
so that the thermodynamic potentials ${\cal S}(E)$ (entropy) and ${\cal J}(\beta)$ (free energy) are
\begin{eqnarray}
{\cal S}(E)&=&-\frac{2}{3}\Omega^2-2E \pm 4\Omega\sqrt{\frac{E}{3}},\\
{\cal J}(\beta)&=&\frac{2}{3} \Omega^2 \frac{\beta}{\beta_1-\beta}.
\end{eqnarray} 

\begin{figure}
\begin{centering}
\includegraphics[width=0.7\textwidth]{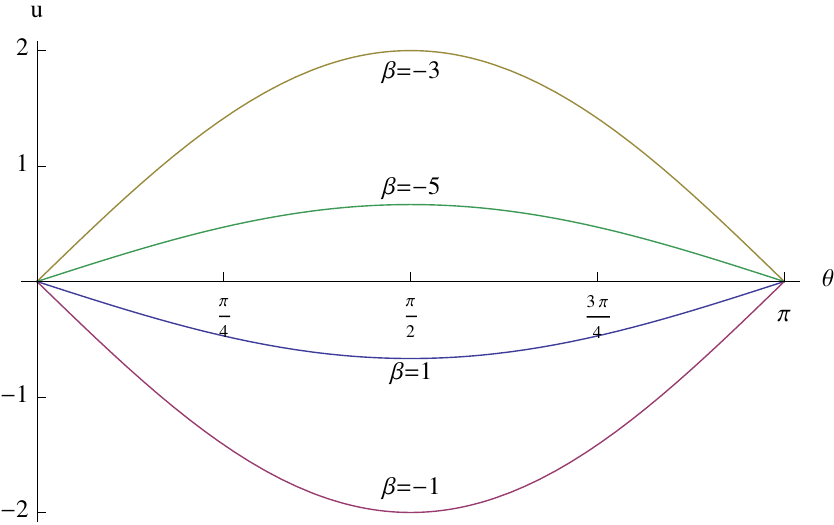}
\caption{Zonal wind profiles for the statistical equilibrium with conservation of energy and circulation for different values of the statistical temperature. The equilibrium flow is a solid body rotation. The zonal wind is antisymmetric with respect to the transformation $\beta \to 2\beta_1 -\beta $. The zonal wind $u$ is  normalized by the choice $R_T=\Omega=1$.}\label{zonalwindEGfig}
\end{centering}
\end{figure}

Knowing the equilibrium streamfunction $\psi$, we can compute the zonal and meridional components of the velocity field, respectively $u$ and $v$:
\begin{equation}
u=-\frac{1}{R_T}\frac{\partial \psi}{\partial \theta}=-\frac{2\Omega}{R_T}\frac{\sin \theta}{\beta-\beta_1},  \quad v=\frac{1}{R_T\sin \theta}\frac{\partial \psi}{\partial \phi}=0,
\end{equation}
where $R_T$ is the Earth's mean radius. The equilibrium motion of the fluid is thus a simple solid body rotation with angular velocity 
\begin{equation}
\Omega_*=-\frac{2\Omega}{R_T^2(\beta-\beta_1)}.
\end{equation}
In particular, the equilibrium velocity distribution is purely zonal, vanishing at the poles with a maximum at the equator. At low statistical temperatures ($\beta < \beta_1=-2$), the rotation of the fluid has the same sign as the solid body rotation of the Earth, while high statistical temperatures ($\beta > \beta_1=-2$) correspond to counter-rotating flows. Examples of such zonal wind profiles are drawn on figure \ref{zonalwindEGfig} for various statistical temperatures $\beta$. Note that for any given value of the energy, the two types of solutions coexist, due to the symmetry $E(-\beta-4)=E(\beta)$, as is clear from the caloric curve $\beta(E)$ shown in figure \ref{caloriccurveEGfig}.

{\it Remark:} when $\beta=0$ (corresponding to $E=\Omega^2/3$), we find that $\Omega_*=-\Omega$ so that there is no 
rotation in the inertial frame. In a sense, the Earth rotation is canceled.

\subsubsection{Case $\beta \in \mbox{ Sp } \Delta$}\label{degenerateHelmholtzEGsection}

Let us now suppose that $\beta = \beta_n$ with $n \neq 1$. The
solutions of equation (\ref{stateHelmholtzEq}) form a $2n+1$
dimensional affine space: if $\psi_0$ is such that $A_\beta \psi_0 =f$
then the space of solutions is simply $\psi_0 + \ker
A_\beta$. Specifically, the general solution reads
\begin{equation}
\psi = \frac{f}{\beta_1-\beta_n}+ \sum_{m=-n}^n \psi_{nm} Y_{nm}(\theta,\phi),
\end{equation}
where $\psi_{nm}$ are arbitrary coefficients, constrained only by the
fixed value of the energy. Clearly, the expressions for the energy and
the entropy become
\begin{eqnarray}\label{degenerateenergyEGeq}
E &=& \frac{4\Omega^2}{3(\beta_n-\beta_1)^2} -\frac{\beta_n}{2} \sum_{m=-n}^n \psi_{nm}^2, \\
S &=& - \frac{2 \Omega^2 \beta_n^2}{3(\beta_n-\beta_1)^2}-\frac{\beta_n^2}{2}\sum_{m=-n}^n \psi_{nm}^2,
\end{eqnarray}
so that the thermodynamic potentials are given by
\begin{eqnarray}
{\cal S}(E)=\beta_n E -\frac{2}{3} \Omega^2 \frac{\beta_n}{\beta_n-\beta_1},\\
{\cal J}(\beta=\beta_n)= \frac{2}{3} \Omega^2 \frac{\beta_n}{\beta_1-\beta_n}.
\end{eqnarray}

For a fixed value of $\beta=\beta_n$, equation (\ref{degenerateenergyEGeq})
means that the energy can have any value greater than
$E(\beta_n)=4\Omega^2/(3(\beta_n-\beta_1)^2)$, depending on the
coefficients $\psi_{nm}$. This degeneracy is apparent in figure
\ref{caloriccurveEGfig}: each time the Lagrange multiplier $\beta$
reaches an eigenvalue of the Laplacian, we have a plateau of the
caloric curve. The degeneracy is in fact multiple: for each point of
the plateau, characterized by $(\beta_n,E_n)$, we have a whole $2n$
dimensional sphere of solutions, with radius
$\sqrt{2(E_n-E(\beta_n))/(-\beta_n)}$.

In the grand-canonical ensemble, the grand-potential has the same
value for all the states on the plateau $\beta=\beta_n$. 

Strictly
speaking, $\beta=\beta_1$ is a forbidden value since the solution
space is then empty. Nevertheless, one can consider that as $\beta \to
\beta_1$, the streamfunction diverges proportionally to $\psi_0$:
\begin{equation}
\psi \sim - \frac{f}{\beta-\beta_1}.
\end{equation}
Similarly, the energy diverges as $(\beta-\beta_1)^{-2}$.

\subsubsection{Nature and stability of the critical points}\label{stabilityEGsection}

So far, we have only found the critical points of the variational
problems (\ref{minensvarprob1}) and (\ref{minensvarprob2}). It remains
to determine their nature: minimum, maximum or saddle points of the
entropy functional. To that purpose, we introduce the grand-potential
functional $J = S - \beta E - \alpha \Gamma$.  A critical point of
entropy at fixed energy and circulation is a local maximum if, and only,
if
\begin{equation}
\delta^2 J = - \int_D \frac{\left( \delta q \right)^2}{2} d^2{\bf r} - \frac{\beta}{2} \int_D \left({\bf \nabla} \delta \psi \right)^2 d^2{\bf r}<0,
\end{equation}
for all perturbations $\delta q$ that conserve energy and circulation
at first order. This is the stability condition in the microcanonical
ensemble. In the grand-canonical ensemble, the stability condition
becomes $\delta^2 J<0$ for all perturbations $\delta q$
\cite{Chavanis2009}.

Clearly if $\beta>0$, $\delta^2 J < 0$ and the point is a maximum of
$S$ with respect to perturbations conserving the energy and
circulation. Actually, this remains true as long as $\beta > \beta_1$
(see \cite{Robert1991a}). In fact, $\delta^2 J < 0$ for all
perturbations $\delta q$, even those which break the conservation of
the constraints: the flow is grand-canonically stable (which implies
microcanonical stability). This is related to the
Arnold sufficient condition of nonlinear dynamical stability
\cite{Chavanis2009}.

Conversely, for $\beta < \beta_1$, let us show that the statistical
equilibria computed in this section are in fact saddle points of the
entropy. It suffices to consider perturbations proportional to
eigenvectors of the Laplacian. Let $\delta \psi_{nm} = \epsilon
Y_{nm}$ such that $\Delta \delta \psi_{nm} = \beta_n \delta
\psi_{nm}$. This perturbation conserves the circulation, and the
variation of the energy at first order is $\delta E = - \epsilon
\beta_n \int_D \psi Y_{nm}\, d^2{\bf r}$ where $\psi$ is the stream
function of the basic mean flow (hence a linear combination of
$Y_{00}$ and $Y_{10}$, and possibly $Y_{pm}$, $-p \leq m \leq p$, $p
\geq 2$ in the degenerate case). By the orthogonality property of the
spherical harmonics, $\delta E=0$ if $n\geq 2$ ($n\neq p$) or $n=1$
and $m\neq 0$. For this particular perturbation $\delta\psi_{nm}$, we
have
\begin{equation}
\delta^2 J = \beta_n (\beta-\beta_n)\int_D \frac{(\delta \psi_{nm})^2}{2} d^2 {\bf r},
\end{equation}
so that $\delta^2 J > 0$ if $\beta<\beta_n$.  Hence, perturbations
proportional to eigenvectors of the Laplacian of order $n$ with $\beta
< \beta_n$ suffice to destabilize the flow. The particular
perturbation $\delta \psi_{11}$ for instance destabilizes the
equilibrium mean flow as soon as $\beta < \beta_1$. In particular, no
degenerate mode is stable. We have proven here microcanonical
instability, which implies grand-canonical instability.  As a
consequence, for any given energy, there is only one stable
equilibrium state (solid line on figure \ref{caloriccurveEGfig} which
corresponds to $\beta>-2$). The dashed lines on figure
\ref{caloriccurveEGfig} correspond to unstable saddle points.

\subsubsection{Summary of the results}

\begin{figure}
\begin{centering}
\includegraphics[width=0.7\textwidth]{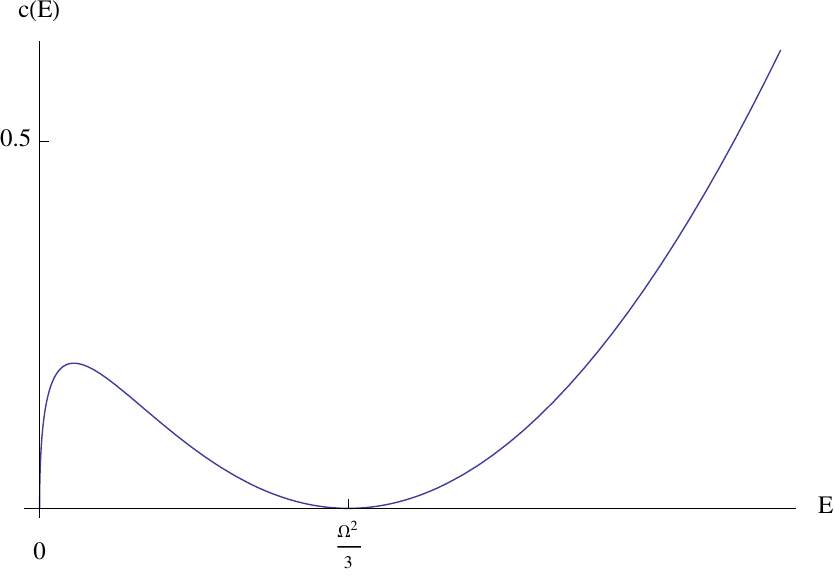}
\caption{Heat capacity $(\partial T/\partial E)^{-1}$ in the case where only the energy and circulation are conserved. The heat capacity vanishes when $E=0$ or $E=\Omega^2/3$, which is also a point of discontinuity of the microcanonical temperature $T=1/\beta$. When $E<\Omega^2/3$, the temperature is positive, otherwise it is negative.}\label{heatcapacityEGfig}
\end{centering}
\end{figure}

In the microcanonical ensemble, there is only one equilibrium state
(global entropy maximum) for each energy. It corresponds to the solid
line on figure \ref{caloriccurveEGfig}. The associated equilibrium
flow is a counter-rotating solid-body rotation. The other states are
unstable saddle points. In the canonical ensemble, there is an
equilibrium state only for $\beta>\beta_1$.The other states are
unstable saddle points. The ensembles are equivalent. The statistical
temperature $T$ is given by
\begin{equation}
\frac{1}{T}=\beta=\frac{\partial {\cal S}}{\partial E}=\sqrt{\frac{4\Omega^2}{3E}}-2.
\end{equation}
It is negative when $E>\Omega^2/3$. The second derivative of the
entropy is negative:
\begin{equation}
\frac{\partial^2 {\cal S}}{\partial E^2} = - \frac{| \Omega |}{\sqrt{3E^3}} \leq 0,
\end{equation}
which means that ${\cal S}(E)$ is a concave function, in accordance
with our findings of ensemble equivalence. Furthermore, the heat
capacity $c=(\partial T/\partial E)^{-1}=(-T^2\partial^2 {\cal
S}/\partial E^2)^{-1}$ is positive and can be computed explicitly (see
figure \ref{heatcapacityEGfig}):
\begin{equation}
c=4\sqrt{\frac{\Omega^2}{3}}E^{1/2}-8E+4\sqrt{\frac{3}{\Omega^2}}E^{3/2}.
\end{equation}

\subsection{Fixed energy, circulation and angular momentum}\label{EGLsection}

Due to the axial symmetry, there is another relevant conserved
quantity that can be taken into account in the variational problem:
the angular momentum
\begin{equation}
L=\langle (q-f) \cos \theta \rangle = \sqrt{\frac{4\pi}{3}} \langle (q-f) | Y_{10}\rangle.
\end{equation}
The critical points of the variational problems
\begin{eqnarray}
{\cal S}(E,\Gamma,L) = \max_q \{ S[q]  | E[q]=E, \Gamma[q]=\Gamma, L[q]=L \},
\end{eqnarray}
and
\begin{eqnarray}
{\cal J}(\beta,\alpha,\mu) = \max_q \{ S[q]  - \beta E[q] - \alpha \Gamma[q] - \mu L[q]\},
\end{eqnarray}
now satisfy
\begin{equation}
\delta S-\beta \delta E - \alpha \delta \Gamma -  \mu \delta L=0,
\end{equation}
which leads to the $q-\psi$ relationship $q=-\beta \psi -\alpha - \mu
\cos \theta$. According to section \ref{qgeqssection}, the solutions
of this equation correspond to states that are steady in a frame
rotating with angular velocity $\Omega_L=\mu/\beta$: indeed, this relation is
of the form $q=F\left(\psi+\frac{\mu}{\beta}\cos\theta\right)$ with
$F(x)=-\beta x-\alpha$.  Imposing $\mu=0$, that is neglecting
conservation of the angular momentum, amounts to considering only the
solutions which are stationary in the reference frame rotating with
the angular velocity of the Earth. These solutions were described in
the previous section. As in the previous section, spatial averaging
yields $\alpha=-\beta
\langle \psi \rangle$ and the $q-\psi$ relationship becomes
\begin{equation}
q = -\beta \left( \psi - \langle \psi \rangle \right) -  \mu \cos \theta.
\end{equation}
Making again the gauge choice $\langle \psi\rangle=0$, leading to $\alpha=0$, and setting
$\tilde{f}=f+ \mu \cos \theta=(2\Omega+ \mu)\cos \theta$, we find that
$\psi$ is given by the Helmholtz equation
\begin{equation} \label{stateHelmholtzEq2}
A_\beta \psi = \tilde{f}.
\end{equation}

We now discuss the resolution of the Helmholtz equation
(\ref{stateHelmholtzEq2}) as in the previous section.

\subsubsection{Case $\beta \notin \mbox { Sp } \Delta$: the continuum solution}

In this case, $A_\beta$ is invertible and $\psi$ is proportional to
the first eigenmode of the Laplacian, so that
\begin{equation}
\psi =  -\frac{\tilde{f}}{\beta-\beta_1}=\frac{2\Omega+ \mu}{\beta_1-\beta}\cos \theta.
\end{equation}
The equilibrium flow is a solid-body rotation with angular velocity
\begin{equation}
\Omega_*=\frac{2\Omega+\mu}{\beta_1-\beta}.
\end{equation}
The potential vorticity is $q=2(\Omega+\Omega_*)\cos \theta$. We can
compute the energy, angular momentum, and entropy:
\begin{eqnarray}
E&=&-\frac{\beta_1}{2} {\Omega_*}^2 \langle \cos^2 \theta\rangle=\frac{1}{3} \left(\frac{2\Omega+\mu}{\beta-\beta_1}\right)^2,\\
L&=&\frac{2}{3}\Omega_*=\frac{2}{3}\frac{2\Omega+\mu}{\beta_1-\beta},\\
S&=&-\frac{2}{3} \left(\Omega+\Omega_*\right)^2=-\frac{2}{3}\left(\frac{\mu-\beta \Omega}{\beta_1-\beta}\right)^2.
\end{eqnarray}
The thermodynamic potentials ${\cal S}(E,L)$ and ${\cal J}(\beta,
\mu)$ are given by
\begin{eqnarray}
{\cal S}(E,L)&=&-\frac{2}{3}\Omega^2-2\Omega L - 2E=-\frac{3}{2}\left(L + \frac{2}{3}\Omega  \right)^2,\\
{\cal J}(\beta, \mu)&=&\frac{1}{3} \frac{(2\Omega+\mu)^2}{\beta-\beta_1}-\frac{2}{3}\Omega^2.
\end{eqnarray}
As always true for solid-body rotations (see
\ref{ELappendix}), the energy and angular momentum are linked by
$E=E^*(L)$, with $E^*(L)=3L^2/4$. This relation is independent of
$\beta$. Hence in the microcanonical ensemble, the continuum solution
exists only on the curve $E=E^*(L)$. For $E>E^*(L)$ there is no such
solution.

\subsubsection{Case $\beta \in \mbox{ Sp } \Delta$}

Let us suppose that $\beta=\beta_n$ with $n \neq 1$.  As in section
\ref{degenerateHelmholtzEGsection}, the general solution of the
Helmholtz equation (\ref{stateHelmholtzEq2}) when $\beta$ is an
eigenvalue of the Laplacian is a superposition of eigenmodes given by
\begin{eqnarray}
\psi &=& -\frac{\tilde{f}}{\beta_n-\beta_1}+ \sum_{m=-n}^n \psi_{nm} Y_{nm}(\theta,\phi)\\
&=& \Omega_n^* \cos \theta + \sum_{m=-n}^n \psi_{nm} Y_{nm}(\theta,\phi),
\end{eqnarray}
where $ \Omega_n^* = \frac{2\Omega+\mu}{\beta_1-\beta_n}$ and
$\psi_{nm}$ are arbitrary coefficients. The requirement for the stream
function to be real-valued imposes $\psi_{n,-m}=\psi_{nm}^*$. The
corresponding energy, angular momentum and entropy are given by
\begin{eqnarray}
E &=& \frac{{\Omega_n^*}^2}{3}  - \frac{\beta_n}{2} \sum_{m=-n}^n |\psi_{nm}|^2, \\
L&=&\frac{2}{3}\Omega_n^*,\\
S &=& -\frac{2}{3}\left(\Omega+\Omega_n^*\right)^2 -\frac{\beta_n^2}{2} \sum_{m=-n}^n |\psi_{nm}|^2.
\end{eqnarray}
The Lagrange multiplier $\mu$, associated with the conservation of
angular momentum, is determined by the relation $L=2\Omega_n^*/3$
which can be inverted to yield
\begin{equation}
\mu=\frac{3}{2} (\beta_1-\beta_n) L - 2\Omega.
\end{equation}
The entropy ${\cal S}(E,L)$ and grand-potential ${\cal J}(\beta,\mu)$
are given by
\begin{eqnarray}
{\cal S}(E,L)=\beta_n\left(E-E^*(L)\right)-\frac{3}{2}\left( L+\frac{2}{3}\Omega\right)^2,\\
{\cal J}(\beta=\beta_n,\mu)=\frac{1}{3} \frac{(2\Omega+\mu)^2}{\beta_n-\beta_1}-\frac{2}{3}\Omega^2.
\end{eqnarray}
We shall see that these solutions are unstable saddle points in both
ensembles.  In the microcanonical ensemble, when $E=E^*(L)$, this
degenerate solution reduces to the continuum solution.

\subsubsection{Case $\beta=\beta_1$}\label{mixedLflowsec}

\begin{figure}
\begin{centering}
\includegraphics[width=\textwidth]{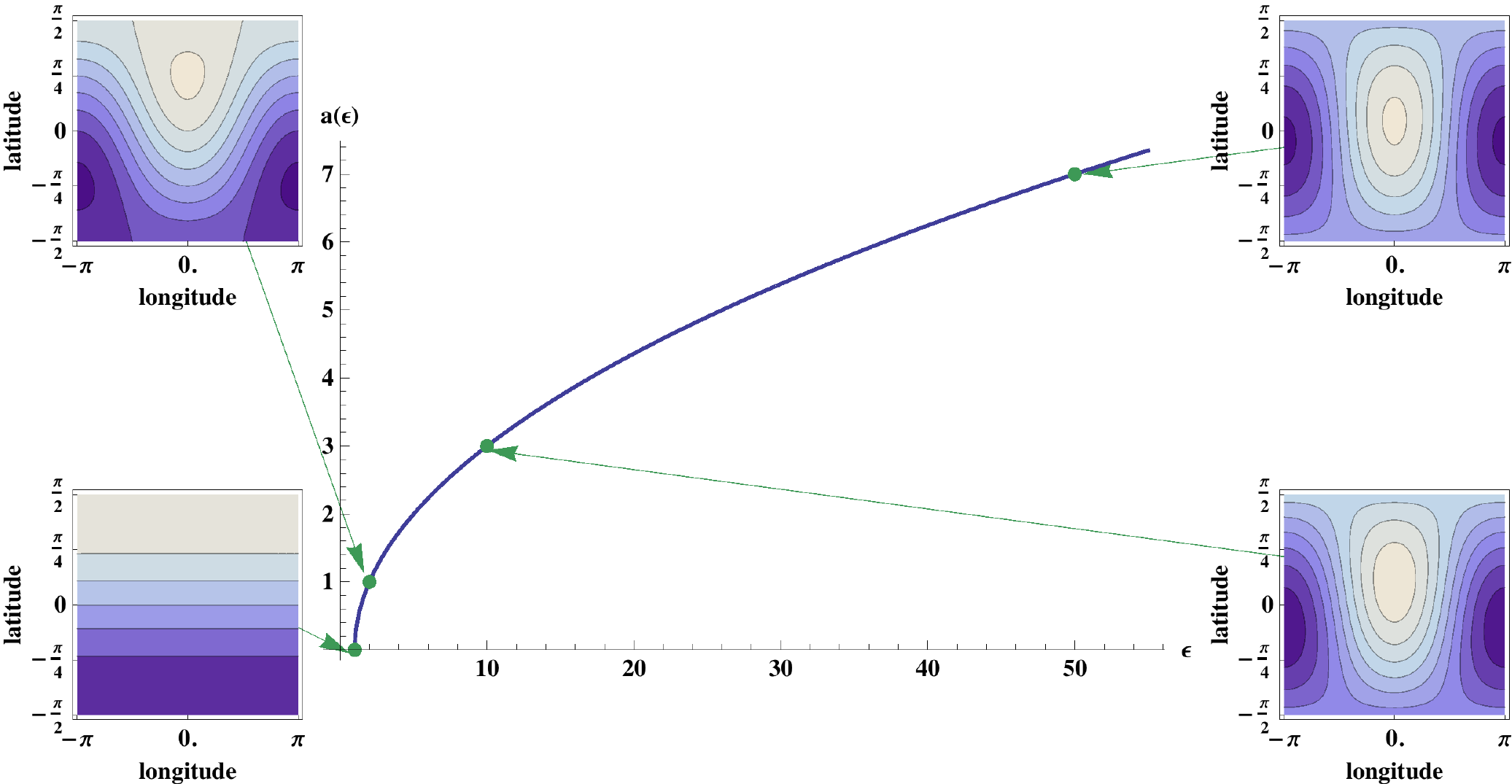}
\caption{Amplitude of the dipole as a function of the control parameter $\epsilon \equiv E/E^*(L)$. There is a second order phase transition at $\epsilon_c=1$ between a ``solid-body rotation" phase ($\epsilon=\epsilon_c$) and a ``dipole" phase ($\epsilon > \epsilon_c$). Insets show particular stream functions for specific values of $\epsilon$. Here $\phi_0=0$.}\label{phasetransitionEGLfig}
\end{centering}
\end{figure}

In this case, equation (\ref{stateHelmholtzEq2}) admits solutions only
if the right-hand side vanishes, i.e. when $\mu=\mu_c \equiv
-2\Omega$. Then, the equilibrium flow has the general form
\begin{equation}
\psi = \psi_{10} Y_{10}(\theta,\phi) + \psi_{11} Y_{11}(\theta,\phi) + \psi_{11}^* Y_{1,-1}(\theta,\phi),
\end{equation}
where $\psi_{10}$ is a real coefficient and $\psi_{11}$ a complex
coefficient, linked by the energy and angular momentum requirements.
Setting $\Omega_*=\sqrt{\frac{3}{4\pi}} \psi_{10}, \gamma_c=\sqrt{\frac{3}{2\pi}} \Re
\psi_{11}, \gamma_s = -\sqrt{\frac{3}{2\pi}} \Im \psi_{11}$, the energy,
angular momentum and entropy read
\begin{eqnarray}
E &=& \frac{1}{3}\left( \Omega_*^2+\gamma_c^2+\gamma_s^2\right),\\
L &=& \frac{2}{3} \Omega_*,\\
S &=&-\frac{2}{3}\left( (\Omega+\Omega_*)^2+\gamma_c^2+\gamma_s^2\right),
\end{eqnarray}
so that $\Omega_*$ is in fact fixed by the angular 
momentum $L$ while $\gamma_c$ and $ \gamma_s$ depend on both $E$ and $L$:
\begin{eqnarray}
\Omega_*&=&\frac{3}{2} L,
\end{eqnarray}
\begin{eqnarray}
\gamma_c^2+\gamma_s^2&=&3\left(E-E^*(L)\right).
\end{eqnarray}
Introducing the angle $\phi_0$ such that
$\gamma_c=\sqrt{3(E-E^*(L))}\cos \phi_0$ and
$\gamma_s=\sqrt{3(E-E^*(L))} \sin \phi_0$, the stream function reads
\begin{eqnarray}
\psi &=& \Omega_* \cos \theta + \gamma_c \sin \theta \cos \phi + \gamma_s \sin \theta \sin \phi\nonumber\\
&=&\Omega_* \cos \theta +\sqrt{3\left(E-E^*(L)\right)} \sin \theta \cos (\phi-\phi_0).
\end{eqnarray}
When $E=E^*(L)$, this solution coincides with the continuum solution:
it is a solid-body rotation. When $E>E^*(L)$, the flow has wave-number
one in the longitudinal direction; it is a dipole with the angle
$\phi_0$ playing the role of a phase. The phase $\phi_0$ is arbitrary (it
is not determined by the constraints). The stream function can be re-written as
\begin{equation}
 \psi =\frac{3}{2} L \left\lbrack\cos \theta +\sqrt{\frac{E}{E^*(L)}-1} \sin \theta \cos (\phi-\phi_0)\right\rbrack.
\end{equation}
Therefore, the amplitude of the dipole depends on a single control
parameter $\epsilon\equiv {E}/{E^*(L)}$ and is given by
$a(\epsilon)=(\epsilon-1)^{1/2}$ (on the other hand $\frac{3}{2}L$
just fixes the amplitude). If we interpret $a$ as the order parameter,
this corresponds to a second order phase transition occurring for
$\epsilon\ge \epsilon_c=1$ between a ``solid rotation'' phase and a
``dipole'' phase (see figure \ref{phasetransitionEGLfig}). Sample
stream functions are shown in figure \ref{phasetransitionEGLfig} for
various values of $\epsilon$. The position of the dipole depends on
the value of $\epsilon$: the larger $\epsilon$, the more the dipole is
aligned along the equator.

Note also that the thermodynamic potentials can be expressed as
\begin{equation}
{\cal S}(E,L)=-2\left(E-E^*(L)\right)-\frac{3}{2}\left(L+\frac{2}{3}\Omega\right)^2,
\end{equation}
\begin{equation}
{\cal J}(\beta=\beta_1,\mu=\mu_c)=-\frac{2}{3}\Omega^2.
\end{equation}
These relations have two implications: (i) the solutions with
different $\phi_0$ have the same entropy (which was expected) so they
are statistically equivalent, and (ii) these solutions have a higher
entropy than the solutions with $\beta=\beta_{n>1}$. As a consequence
of (i), the second order phase transition is accompanied by
spontaneous symmetry breaking, as the phase of the dipole is
arbitrary.

{\it Remark:} The condition $\mu=-2\Omega$ with $\beta=\beta_1=-2$ 
corresponds to $\Omega_L=\mu/\beta=\Omega$. 
Therefore, the dipole is stationary in a frame rotating with angular 
velocity $\Omega$ with respect to the Earth (hence, rotating at the angular
 velocity $2\Omega$ with respect to the inertial frame).

\subsubsection{Nature and stability of the critical points}

A critical point of entropy at fixed energy, circulation and angular
momentum is a local maximum if, and only, if
\begin{equation}
\delta^2 J = \delta^2 S - \beta \delta^2 E = - \int_D \frac{\left( \delta q \right)^2}{2} d^2{\bf r} - \frac{\beta}{2} \int_D \left({\bf \nabla} \delta \psi \right)^2 d^2{\bf r}<0,
\end{equation}
for all perturbations $\delta q$ that conserve energy, circulation and
angular momentum at first order. We have introduced the
grand-potential functional $J[q]=S[q]-\beta E[q]-\alpha \Gamma[q] -
\mu L[q]$. This is the stability condition in the microcanonical
ensemble. In the grand-canonical ensemble, the stability condition
becomes $\delta^2 J<0$ for all perturbations $\delta q$
\cite{Chavanis2009}.

Carrying out the same analysis as in section \ref{stabilityEGsection},
one concludes that the critical points found previously are entropy
maxima only if $\beta > \beta_1$. As in section
\ref{stabilityEGsection}, if $\beta > \beta_1$, the flow is
grand-canonically stable (i.e. stable for all perturbations $\delta q$
and not only those which preserve the constraints at first order) and
thus also microcanonically stable.

Otherwise, one can exhibit perturbations that destabilize the mean
flow. Indeed, at first order, perturbations of the type $\delta
\psi_{nm} = \epsilon Y_{nm}$ conserve the circulation as
previously. Since $\delta L = \langle \cos \theta \delta q_{nm}
\rangle$ and $ \cos \theta$ is proportional to $Y_{10}$,
$\delta\psi_{nm}$ conserves the angular momentum if $(n,m)\neq
(1,0)$. Besides, as $\delta E = - \beta_n \langle \psi\delta \psi_{nm}
\rangle$, the perturbation conserves energy for $(n,m)\neq (0,0),
(1,0)$ in the case of the continuum solution, and for $(n,m)\neq
(0,0), (1,0), n\neq p$ when $\beta=\beta_p$. Furthermore, since
$\delta^2 J = \beta_n (\beta-\beta_n)\int_D \frac{(\delta
\psi_{nm})^2}{2} d^2 {\bf r}$, these perturbations destabilize the
mean flow if, and only, if $\beta < \beta_n$. In particular, as soon
as $\beta < \beta_1$, the mean flow is not stable with respect to the
perturbation $\delta\psi_{11}$ for instance. All the equilibrium flows
with $\beta < \beta_1$ are thus saddle points. Again, as in section
\ref{stabilityEGsection}, we have proved microcanonical instability,
which implies grand-canonical instability.

It remains to be seen what happens when $\beta=\beta_1$. In that case,
the quadratic form $\delta^2 J$ is degenerate. The vector space
spanned by $Y_{11},Y_{1,-1}$ and $Y_{10}$ constitutes the radical of
$\delta^2 J$: the function $J$ is constant on this vector space (with
value $-2\Omega^2/3$). Hence we have a three-dimensional vector space
of metastable states in the grand-canonical ensemble. Spontaneous
perturbations may be generated at no cost in inverse temperature
$\beta$ and Lagrange multiplier $\mu$, which induce transitions between
one dipole to another, possibly with different values of energy,
angular momentum, and phase. In the microcanonical ensemble, as we fix
the values of the energy and angular momentum, the only such
perturbation which is possible is that which changes the phase of the
dipole. Hence we only have a one-dimensional manifold of metastable
states in the microcanonical ensemble. These spontaneous perturbations
can be interpreted in terms of Goldstone bosons, as they appear due to
continuous symmetry breaking \cite{ZinnJustinBook}.

\subsubsection{Summary of the results}

\begin{figure}
\begin{centering}
\includegraphics[width=0.48\textwidth]{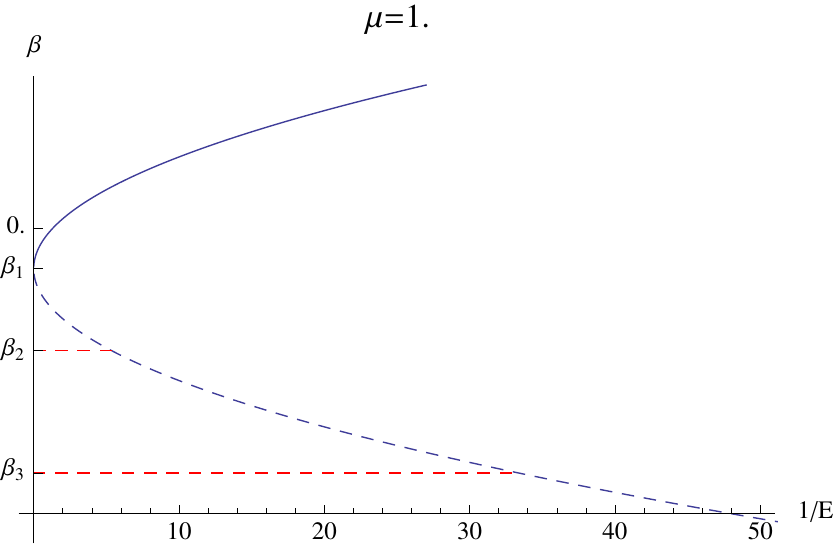}
\includegraphics[width=0.48\textwidth]{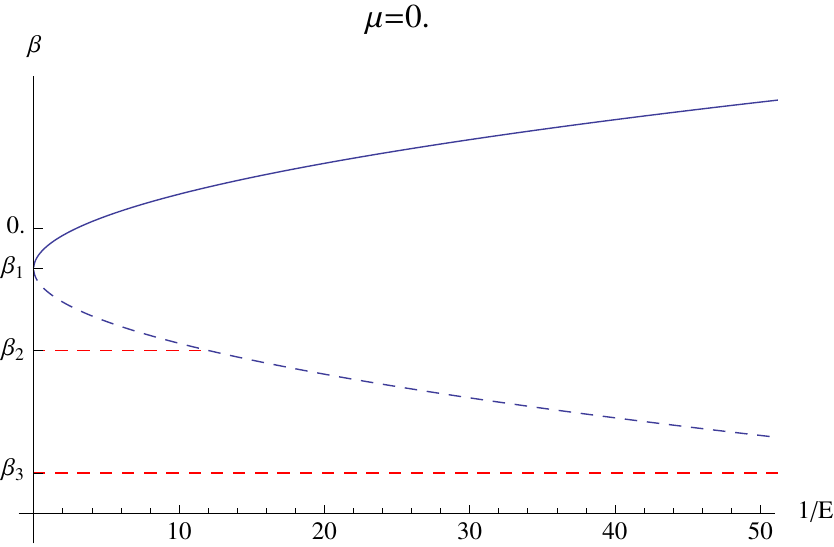}
\includegraphics[width=0.48\textwidth]{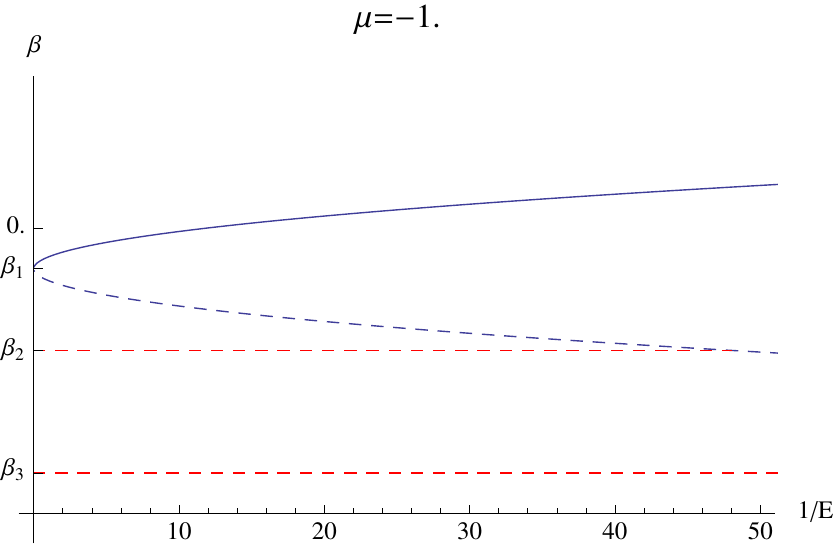}
\includegraphics[width=0.48\textwidth]{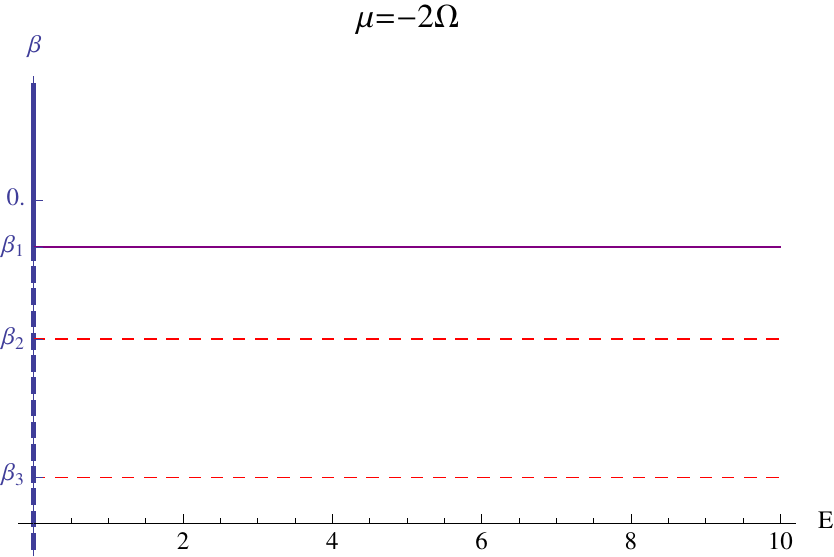}
\caption{Caloric curves $1/E(\beta)$ (respectively $E(\beta)$ for 
the lower-right panel) for different values of the Lagrange parameter $\mu$ in the 
grand-canonical ensemble. From left to right and from top to bottom, $\mu=1,0,-1$ and $\mu=-2\Omega$.
The solid blue line (continuum solution, solid-body rotation) corresponds to true maxima 
of the grand-potential while the dashed blue line corresponds to saddle-points 
(still for the continuum solution). Dashed horizontal red lines indicate the position of the
 eigenvalues of the Laplacian, and therefore correspond to plateaux of degenerate (saddle) solutions.
 In the lower-right panel, $\mu+2\Omega=0$: the continuum solution only exists on the axis $E=0$.
 The solid purple line represents the symmetry-breaking dipole solution.}\label{caloriccurveEGL1fig}
\end{centering}
\end{figure}

\begin{figure}
\begin{center}
\includegraphics[width=0.7\textwidth]{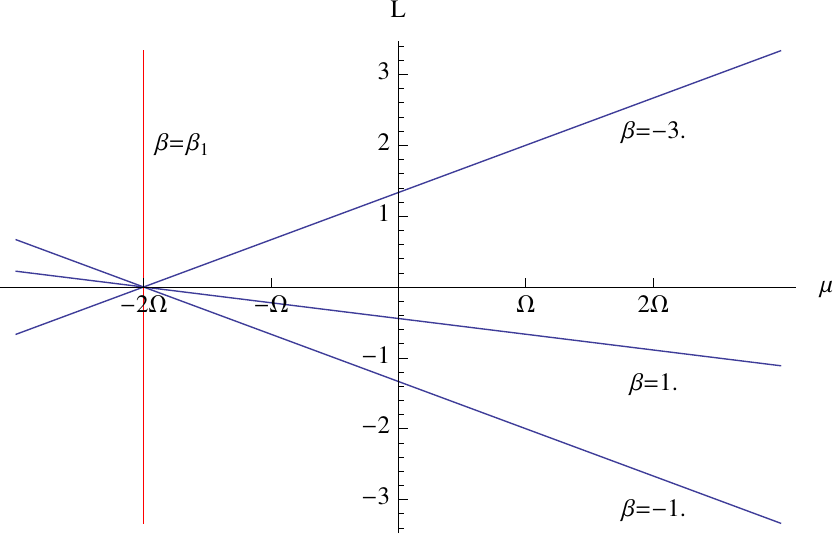}
\caption{Chemical potential curve $L(\mu)$ for different values of the temperature 
$\beta$ in the grand-canonical ensemble. For every value of $\beta$, the curve is a straight line. 
For all $\beta \neq \beta_1$, it has a 
finite slope $2/(3(\beta_1-\beta))$.
 When $\beta=\beta_1$, necessarily $\mu=-2\Omega$, and the value of the angular momentum does not 
depend on $\mu$.}
\label{chempotcurveEGLfig}
\end{center}
\end{figure}

To sum up the results obtained in the previous sections, we start by
treating the grand-canonical ensemble where $\beta$ and $\mu$ are
fixed:
\begin{itemize}
\item If $\mu \neq \mu_c$, the only stable equilibrium state is 
a solid-body rotation $\Omega_*<0$, obtained for $\beta> \beta_1$. Two
types of saddle points are possible for $\beta<\beta_1$: solid-body
rotation $\Omega_*> 0$ when $\beta$ is not an eigenvalue of the
Laplacian, or more structured flows when $\beta$ is an eigenvalue of
the Laplacian but these solutions are unstable. Finally, there is no
solution with $\beta=\beta_1$.

\item If $\mu=\mu_c$, the continuum solution is the trivial motionless 
solution: $\Omega_*=0$ (and thus $E=0, L=0$ for all values of $\beta$,
see figure \ref{caloriccurveEGL1fig}). The eigenmodes solutions remain
accessible but unstable, while a new dipole solution appears for
$\beta=\beta_1$, with arbitrary energy and angular momentum (see
figures \ref{caloriccurveEGL1fig} and \ref{chempotcurveEGLfig}). This
corresponds to a second order phase transition with spontaneous
symmetry-breaking.
\end{itemize}

\begin{figure}
\begin{centering}
\includegraphics[width=0.7\textwidth]{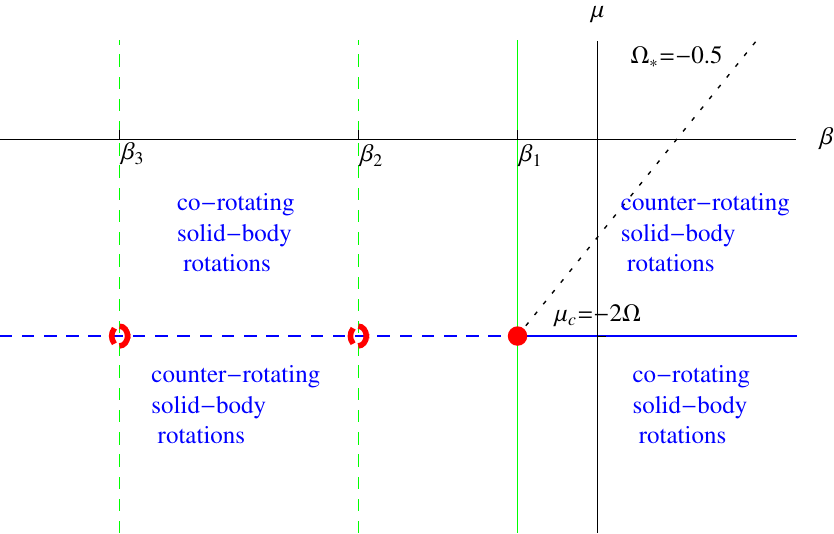}
\caption{Phase diagram in the grand-canonical ensemble. When $\beta>\beta_1$, the equilibrium state is a solid-body rotation, 
co-rotating or counter-rotating depending on the position of $\mu$ with respect to $-2\Omega$. 
The solid blue line indicates the separating case of a motionless flow. When $\beta=\beta_1$ 
and $\mu=-2\Omega$, the equilibrium flow is a symmetry-breaking dipole. There 
is a second order phase transition at this point (red dot). When 
$\beta=\beta_n$, $n\neq 1$ (dashed green lines), solid-body rotations coexist with degenerate 
states, but they are all unstable. Only the degenerate states remain when $\mu=-2\Omega$ (dashed 
red circles). The dashed blue line corresponds to an unstable motionless case, while the 
solid green line is an impossible case (no solution to the mean-field equation). 
The dotted half straight line represents an iso-$\Omega_*$ line (corresponding to 
$\Omega_*=-0.5$).}\label{phasediagramEGLgcfig}
\end{centering}
\end{figure}

For both cases, the caloric curve $E(\beta)$ is shown, for different
values of $\mu$, in figure \ref{caloriccurveEGL1fig}. Similarly, the
curve $L(\mu)$ is shown on figure
\ref{chempotcurveEGLfig} for different values of $\beta$: for $\beta\neq \beta_1$, it is a straight
line with slope $2/(3(\beta_1-\beta))$. For $\beta=\beta_1$, it is a
vertical line at $\mu=-2\Omega$ indicating that the
value of the angular momentum is arbitrary. These results are
summarized in the grand-canonical phase diagram (figure
\ref{phasediagramEGLgcfig}).

\begin{figure}
\begin{center}
\includegraphics[width=0.7\textwidth]{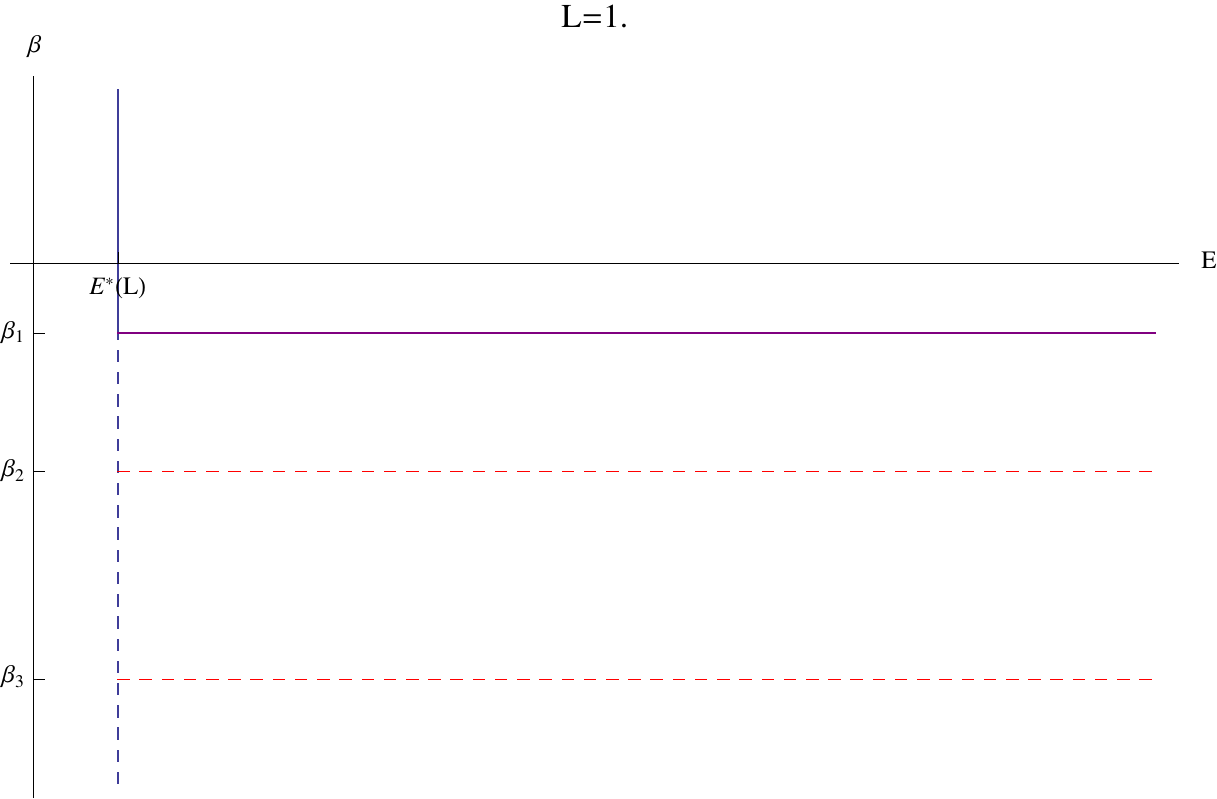}
\caption{Caloric curve $\beta(E)$ in the microcanonical ensemble, in the case when 
the energy, circulation and angular momentum are conserved. For a given value of the
 angular momentum $L$ (here $L=1$), the energy is necessarily greater than $E^*(L)$. 
When $E>E^*(L)$, the only stable equilibrium is obtained for $\beta=\beta_1$ (solid purple 
line, dipole). However, there are an infinity of saddle points corresponding to $\beta=\beta_n$ 
(dashed red lines, degenerate states). When $E=E^*(L)$, $\beta$ is not fixed and can take any 
value. In this case, the flow is a solid-body rotation. Cases $\beta>\beta_1$ correspond to 
stable equilibria while $\beta<\beta_1$ correspond to saddle points. Note that the value of 
the angular momentum $L$ only modifies the position of the point $E^*(L)$, but does not 
change the shape of the microcanonical caloric curve.}\label{microcanonicalcaloriccurvefig}
\end{center}
\end{figure}

\begin{figure}
\begin{center}
\includegraphics[width=0.7\textwidth]{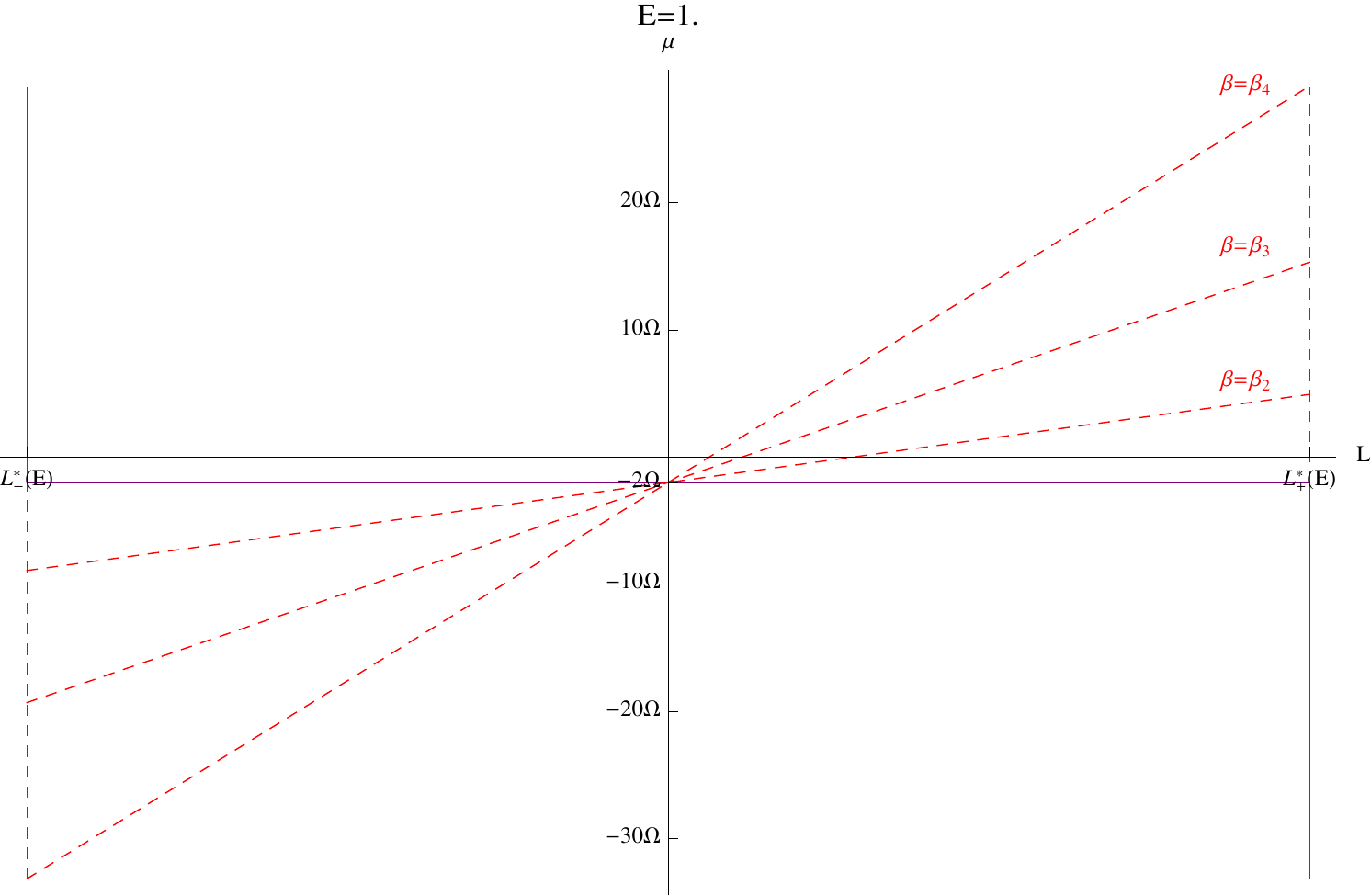}
\caption{Chemical potential $\mu(L)$ in the microcanonical ensemble,
 in the case of conservation of energy, circulation and angular momentum. For a given 
value of $E$ (here $E=1$), $L$ lies between $L_-^*(E)$ and $L_+^*(E)$. The two straight 
lines $L=L_-^*(E)$ and $L=L_+^*(E)$ correspond to solid-body rotations. In this case, 
the value of the parameter $\mu$ is not fixed by $L$ as only the angular velocity $\Omega_*$, 
which is a function of both $\mu$ and $\beta$, is important. The solid blue line represents 
stable solid-body rotations while the dashed blue line corresponds to unstable solid-body 
rotations. The straight line $\mu=-2\Omega$ (solid purple) corresponds to the case of the 
dipole flow, which occurs when $|L|\neq L_+^*(E)$. There are an infinity of straight lines 
with slopes $3(\beta_1-\beta_n)/2$ (three of them are represented with dashed red on the 
figure for $n=2,3,4$), corresponding to unstable degenerate states.}\label{microcanonicalchemicalpotentialfig}
\end{center}
\end{figure}

Now, in the microcanonical ensemble, the equilibrium is determined by
the given value of $(E,L)$ as follows:
\begin{itemize}
\item If $E=E^*(L)$, the stable equilibrium is a solid-body 
rotation with angular velocity $\Omega_*=3L/2$. Note that in this
case, the Lagrange multipliers $\beta$ and $\mu$ are not determined by
$E$ and $L$ (see figures \ref{microcanonicalcaloriccurvefig} and
\ref{microcanonicalchemicalpotentialfig}). The only constraints are
$\beta>\beta_1$ and $\mu<\mu_c$ or $\mu>\mu_c$ depending on the sign
of $L$. In other words, for $E=E^*(L)$, the caloric curve $\beta(E)$
(figure \ref{microcanonicalcaloriccurvefig}) and the chemical
potential line $\mu(L)$ (figure
\ref{microcanonicalchemicalpotentialfig}) are vertical lines.

\item If $E > E^*(L)$, the most probable state is the dipole 
of section \ref{mixedLflowsec}, with an undetermined phase
$\phi_0$. This is a case of spontaneous symmetry breaking, insofar as
the longitude dependence of one particular solution (for a given
$\phi_0$) breaks the axial symmetry. However, as usual, the ensemble of
solutions satisfy the axial symmetry. Furthermore, there are
degenerate solutions which are unstable saddle points with a lower
entropy. The caloric curve $\beta(E)$ at fixed angular momentum
(figure \ref{microcanonicalcaloriccurvefig}) consists of an ensemble
of horizontal lines. The horizontal line with $\beta=\beta_1$
corresponds to the equilibrium dipole flow: the statistical
temperature does not depend on the energy. In addition, there are
horizontal lines at $\beta=\beta_n$, $n>1$ corresponding to the
unstable degenerate states. Similarly, for fixed energy $E$, the
chemical potential curve $\mu(L)$ (figure
\ref{microcanonicalchemicalpotentialfig}) is a horizontal line at
$\mu=\mu_c$ for the (stable) dipole equilibrium. There are also an
infinity of straight lines corresponding to degenerate modes with
$\beta=\beta_n$, $n>1$, with slopes $3(\beta_1-\beta_n)/2$, but these
modes are unstable saddle points.

\end{itemize}

\begin{figure}
\begin{centering}
\includegraphics{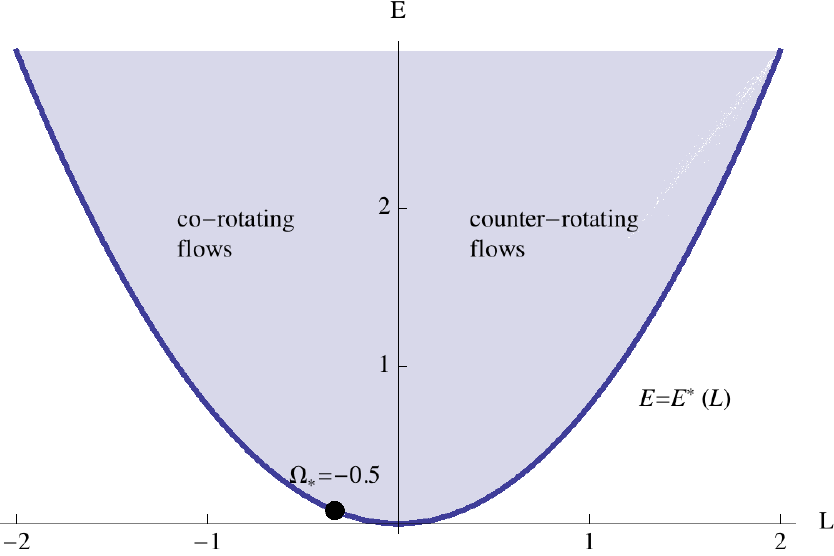}
\caption{Phase diagram in the microcanonical ensemble: the final state of the flow predicted 
by statistical mechanics depends on the position in the $(E,L)$ space. The thick blue line 
represents the curve $E=E^*(L)$ defined in the text. On this curve, the statistical equilibrium 
is a solid-body rotation (counter-rotating for $L>0$ and co-rotating for $L<0$) 
with angular velocity $\Omega_*$ (we have 
shown $\Omega_*=-0.5$). In the portion 
of the plane lying over this curve (blue filled area), the statistical equilibrium is the dipolar 
flow of section \ref{mixedLflowsec}. The blue parabola is the locus of a second order phase 
transition with spontaneous symmetry breaking. The portion under the curve is forbidden by 
the energy inequality obtained in \ref{ELappendix}.}\label{regimediagrammefig}
\end{centering}
\end{figure}

Recall that a flow with energy $E$ and angular momentum $L$
necessarily satisfies $E\geq E^*(L)$ (see \ref{ELappendix}). Therefore
our classification of the final equilibrium state reached by the flow
is complete; it is summarized in the microcanonical phase diagram of
figure \ref{regimediagrammefig}. The line $E=E^*(L)$ corresponds to a
line of second order phase transition with spontaneous symmetry
breaking: on this line, the equilibrium is a solid-body rotation (with
a direction given by the sign of the angular momentum); above the
line, the equilibrium is a dipole flow with amplitude $a(\epsilon)$
and an arbitrary phase.

\subsubsection{Discussion of the ensemble equivalence properties}

Contrary to other studies with the same model (quasi-geostrophic
equations) but in a different geometry
\cite{Venaille2009,Venaille2011}, the microcanonical and the
grand-canonical ensemble are equivalent here. However, the ensemble
equivalence is only \emph{partial} in the standard terminology
\cite{Ellis2000}: we have seen that at the macrostates level, the
equilibrium states reached in the grand-canonical ensemble and in the
microcanonical ensemble are the same. More precisely, each set of
equilibrium states obtained in the microcanonical ensemble (at fixed
$(E,L)$ with $E>E^*(L)$) is a proper subset of the set of
grand-canonical states obtained at the corresponding Lagrange
multiplier $(\beta=\beta_1, \mu=\mu_c)$. This is the general case of
partial ensemble equivalence. Here the situation is rather extreme as
the set of grand-canonical equilibrium states obtained for a single
value of the $(\beta,\mu)$ couple contains all the microcanonical
equilibrium states for any value of the energy and angular
momentum. In other words, the point $(\beta=\beta_1,\mu=\mu_c)$ in the
grand-canonical phase diagram is mapped onto the whole domain
$E>E^*(L)$ in the microcanonical phase diagram. As far as solid-body
rotations are concerned, each half straight line corresponding to a
fixed angular velocity in the grand-canonical phase diagram is mapped
onto a point on the $E=E^*(L)$ parabola in the microcanonical phase
diagram (see figures \ref{phasediagramEGLgcfig} and \ref{regimediagrammefig}).  
Partial ensemble equivalence is also seen at the
thermodynamic level: geometrically, the entropy ${\cal
S}(E,L)=-\frac{2}{3}\Omega^2-2\Omega L-2E$ is a plane. In particular,
it is a concave function, but only marginally so; it is also a convex
function. The statistical temperature $1/T=\beta=\partial {\cal
S}/\partial E$ is constant and equal to $\beta_1$, except possibly
when $E=E^*(L)$. Besides, both second partial derivatives
$\partial^2{\cal S}/\partial E^2$ and $\partial^2{\cal S}/\partial
L^2$ vanish.

Here, it is possible to measure how severe the partial ensemble
equivalence is. We have already described precisely the relationships
between the different sets of equilibrium states obtained for various
values of the parameters in both statistical ensembles. Now, we recall
that in the stability analysis, we mentioned that in the
grand-canonical ensemble, there is metastability in a
three-dimensional vector space, while it reduces to a
one-dimensional manifold in the microcanonical ensemble. In other
words, there are three different modes (Goldstone modes) which can
move the system from one metastable state to another in the
grand-canonical ensemble, while there is only one such Goldstone mode
in the microcanonical ensemble. If we considered any mixed ensemble,
with one constraint treated microcanonically and the other
canonically, we would have two Goldstone modes. Thus, the number of
Goldstone modes in each ensemble provides a refined characterization
of ensemble equivalence properties, as compared to simply calling it
``partial".

The phase transition observed here is made possible by the degeneracy
of the first eigenspace of the Laplacian on the sphere, which allows
for non-trivial energy condensation. Although all the energy condenses
in the first mode, we have two distinct equilibrium flow
structures. The ensemble equivalence properties also owe to the
particular geometry. Previous studies all assumed that the first
eigenvalue of the Laplacian is non-degenerate
\cite{Venaille2009,Venaille2011,Bouchet2010,CorvellecThesis}. When this 
is not the case, it is straightforward to see that ensemble
inequivalence results such as those obtained in \cite{Venaille2009} may
collapse.

Nonetheless, it is not clear how generic partial equivalence of
statistical ensemble is. For instance, considering small
non-linearities in the $q-\psi$ relationship may change the nature of
the phase transition and the ensemble equivalence properties. In the
case of the energy-enstrophy ensemble on a rectangular domain with
fixed boundary conditions, it has been shown \cite{CorvellecThesis}
that the phase transition may remain second order or turn into a
first-order phase transition depending on the sign of the first
non-linear coefficient in the $q-\psi$ relationship. This analysis is
likely to remain valid in the case we are considering here.


\section{General case: quasi-geostrophic flow over a topography.}\label{generalQG1section}

In the previous section, we have seen that in the absence of a bottom
topography, the statistical mechanics of the quasi-geostrophic
equations in spherical geometry can be solved in a very simple manner
due to the fact that the Coriolis parameter is an eigenvector of the
Laplacian on the sphere. Adding an angular momentum conservation
constraint does not alter this derivation since it simply adds another
contribution proportional to the same eigenvector of the Laplacian to
the mean field equation. In this section, we treat the general case,
with a finite Rossby deformation radius as well as an arbitrary
topography.

\subsection{The general mean field equation and its solution}

The critical points of the variational problem
\begin{equation}
{\cal S}(E,\Gamma,L) = \max_q \{ S[q]  | E[q]=E, \Gamma[q]=\Gamma, L[q]=L \},
\end{equation}
given by $\delta S - \beta \delta E - \alpha \delta \Gamma - \mu
\delta L = 0$, satisfy the linear $q-\psi$ relationship $q = - \beta
\psi - \alpha - \mu \cos \theta$. Now, $\alpha$ is determined by
averaging over the whole domain
\begin{equation}
\Gamma=\langle q \rangle = \frac{\langle \psi \rangle}{R^2} + \langle h \rangle = - \beta \langle \psi \rangle - \alpha.
\end{equation}
Replacing $q$ with $-\Delta \psi +\frac{\psi}{R^2}+h$, the mean field
equation becomes
\begin{equation}
A_\lambda \left( \psi - \langle \psi \rangle \right) = h - \langle h \rangle + \mu \cos \theta,
\end{equation}
where $\lambda = \beta + \frac{1}{R^2}$ plays the role of the inverse
temperature $\beta$ in the case of a finite Rossby deformation
radius. As before, we are free to make the gauge choice $\langle \psi
\rangle =0$ as it does not affect the structure of the
flow. Therefore, the mean field equation can be rewritten
\begin{equation}
-\Delta \psi +\lambda \psi= \langle h \rangle -h-\mu \cos \theta.
\label{mow}
\end{equation}
Following \cite{Naso2011}, the general solution can be written
\begin{equation}\label{psiHelmoltzsol}
\psi = \phi_1 +\mu \phi_2,
\end{equation}
where $\phi_1$ and $\phi_2$ satisfy
\begin{eqnarray}
-\Delta \phi_1 +\lambda \phi_1&=&\langle h \rangle -h,\\
-\Delta \phi_2 +\lambda \phi_2&=& - \cos \theta.
\end{eqnarray}

Let us assume for the moment that $\lambda \notin \mbox{Sp } \Delta$. We have
\begin{equation}\label{phi2series}
\phi_1 = - \sum_{n \neq 0} \sum_{m=-n}^n \frac{\langle h | Y_{nm}\rangle}{\lambda-\beta_n} Y_{nm},
\end{equation}
and
\begin{equation}
\phi_2 = - \sqrt{\frac{4\pi}{3}} \frac{1}{\lambda-\beta_1} Y_{10}.
\end{equation}
The general solution reads
\begin{equation}\label{psigeneral}
\psi = -  \sqrt{\frac{4\pi}{3}} \frac{\mu}{\lambda-\beta_1} Y_{10} + \sum_{n=1}^{+\infty} \sum_{m=-n}^n \frac{\langle h | Y_{nm} \rangle}{\beta_n - \lambda} Y_{nm}(\theta, \phi),
\end{equation}
or in terms of the potential vorticity
\begin{equation}
q = \langle h \rangle + \sqrt{\frac{4\pi}{3}} \frac{\beta_1-R^{-2}}{\lambda-\beta_1} \mu Y_{10} - \beta \sum_{n=1}^{+\infty} \sum_{m=-n}^n \frac{\langle h | Y_{nm} \rangle}{\beta_n - \lambda} Y_{nm}(\theta, \phi).
\end{equation}
Defining $\tilde{h}=h+\sqrt{\frac{4\pi}{3}}\mu Y_{10}$, the stream function
takes the compact form
\begin{eqnarray}
\psi &=& \sum_{n=1}^{+\infty} \sum_{m=-n}^n \frac{\langle \tilde{h} | Y_{nm} \rangle}{\beta_n - \lambda} Y_{nm}.
\end{eqnarray}
The energy is given by the relation
\begin{equation}\label{Edefeqn}
E = - \frac{1}{2} \langle \psi \Delta \psi \rangle + \frac{1}{2} \frac{\langle \psi^2 \rangle}{R^2},
\end{equation}
which gives, after replacing with equations (\ref{mow}) and (\ref{psiHelmoltzsol}), and simplifying:
\begin{equation}\label{energymideqn}
E =\frac{\langle \phi_1 | \phi_1 \rangle}{8\pi} \left( R^{-2}-\beta\right)-\frac{\langle h | \phi_1\rangle}{8\pi} + \frac{2R^{-2}-\beta_1}{(\lambda-\beta_1)^2}\mu\left(\frac{\mu}{6}+ \frac{\langle h | Y_{10} \rangle }{\sqrt{12\pi}}\right).
\end{equation}
Replacing $\phi_1$ with equation (\ref{phi2series}) - or directly substituting equation
(\ref{psigeneral}) in equation (\ref{Edefeqn}) - we obtain $E$ as the sum of a
series:
\begin{equation}
E =  \frac{1}{8\pi}\sum_{n=1}^{+\infty}\sum_{m=-n}^n  \frac{|\langle \tilde{h} | Y_{nm} \rangle|^2}{(\beta_n-\lambda)^2}\left( R^{-2}-\beta_n\right).\label{energygeneraleqn}
\end{equation}
Similarly, the angular momentum and entropy can be expressed as
\begin{eqnarray}
\fl L =  -\beta \frac{\langle \psi | Y_{10} \rangle}{\sqrt{12\pi}} - \frac{\mu}{3}
 = \frac{1}{3} \frac{\beta_1-R^{-2}}{\lambda-\beta_1}\mu + \frac{\beta}{\lambda-\beta_1}\frac{\langle h | Y_{10} \rangle}{\sqrt{12\pi}} ,\label{angmomgeneraleqn} \\
\fl
S = - \frac{1}{2} \langle h \rangle^2 - \frac{\left(\sqrt{4\pi/3}\mu(\beta_1-R^{-2})+\beta\langle h | Y_{10}\rangle\right)^2}{8\pi(\beta_1-\lambda)^2}  -  \frac{\beta^2}{8\pi} \sum_{n,m} \frac{|\langle h |Y_{nm} \rangle|^2}{(\beta_n-\lambda)^2},\label{entropygeneraleqn}
\end{eqnarray}
where the sum is on all indices $(n,m)$ except $(0,0)$ and $(1,0)$.

This is the general solution of the problem. For a given topography,
these equations of state determine the Lagrange multipliers $\beta$ and
$\mu$ as a function of the energy $E$ and angular momentum
$L$. Equation (\ref{angmomgeneraleqn}) is easily inverted to yield
\begin{equation}\label{muofL}
\mu = \frac{3\beta}{R^{-2}-\beta_1} \frac{\langle h | Y_{10} \rangle}{\sqrt{12\pi}}  -\frac{3(\lambda-\beta_1)}{R^{-2}-\beta_1}L.
\end{equation}
Similarly, the relation between the Lagrange multiplier $\alpha$ and
the circulation $\Gamma$ is easily obtained:
\begin{equation}\label{alphaofgammaeq}
\alpha=-\beta \langle \psi \rangle - \Gamma=-\Gamma=-\langle h\rangle.
\end{equation}
From equations (\ref{energymideqn}) and (\ref{angmomgeneraleqn}), we introduce
the control parameter 
\begin{equation}
{\cal E}(E,L)\equiv  E + 3 \frac{2R^{-2}-\beta_1}{(R^{-2}-\beta_1)^2} L  \left(\frac{\langle h | Y_{10} \rangle}{\sqrt{12\pi}} -\frac{L}{2} \right),
\end{equation}
in terms of which we obtain
\begin{eqnarray}\label{Ectrlfunbeta}
{\cal E}=\frac{\langle \phi_1 | \phi_1 \rangle}{8\pi} \left( R^{-2}-\beta\right)
-\frac{\langle h | \phi_1\rangle}{8\pi} \nonumber\\
+ \frac{2R^{-2}-\beta_1}{(R^{-2}-\beta_1)^2}\frac{\beta(\lambda 
+R^{-2}-2\beta_1)}{(\lambda-\beta_1)^2} \frac{\langle h | Y_{10} \rangle ^2}{8\pi}.
\end{eqnarray}
The right hand-side is only a function of $\beta$. Thus, in practice,
for a given set of control parameters $({\cal E},L,\Gamma)$, solving
equation (\ref{Ectrlfunbeta}) for $\beta$ together with equations
(\ref{muofL}) and (\ref{alphaofgammaeq}) yields the value of the
Lagrange multipliers $(\beta,\mu,\alpha)$.

If we do not take into account the conservation of angular momentum
($\mu=0$), equations (\ref{energymideqn}) and (\ref{angmomgeneraleqn})
reduce to
\begin{eqnarray}\label{Emuzeroeq}
E&=&\frac{\langle \phi_1 | \phi_1 \rangle}{8\pi} \left( R^{-2}-\beta\right)-\frac{\langle h | \phi_1\rangle}{8\pi},\\
L&=&\frac{\beta}{\lambda-\beta_1}\frac{\langle h | Y_{10} \rangle}{\sqrt{12\pi}}.
\end{eqnarray}
Since $\langle \phi_1 | \phi_1 \rangle=\frac{\partial \langle h |
\phi_1 \rangle}{\partial \lambda}$, equation (\ref{Emuzeroeq}) can
also be written
\begin{equation}
E=\frac{1}{8\pi} \frac{\partial}{\partial \lambda}\left( (2R^{-2}-\lambda)\langle h | \phi_1 \rangle\right).
\end{equation}
In the case $R=\infty$, this expression reduces to $E=-\frac{1}{2}
\frac{d}{d\beta}\left(\beta\langle h \phi_1 \rangle\right)$ as found
in \cite{Naso2011}.

\subsection{Degenerate solutions of the mean field equation}

In the case where $\lambda \in \mbox{Sp } \Delta$, if $\lambda =
\beta_p$ with $p> 1$, then, strictly speaking, we must have $\langle h
Y_{pm} \rangle = 0$ for all $-p \leq m \leq p$. In practice, if the
topography has a non-vanishing mode of order $(p,m)$, which is the
generic case, then as $\lambda \to \beta_p$, the corresponding mode
becomes overwhelmingly dominant in the stream function: $\psi \sim
\frac{\langle h | Y_{pm} \rangle}{\beta_p - \lambda} Y_{pm}$ (if there
are several possible values of $m$, then the stream function will be
proportional to the appropriate linear combination of these
modes). The corresponding divergence for the energy and entropy are
like $(\lambda-\beta_p)^{-2}$.  If, on the contrary, the topography
has no contribution proportional to any spherical harmonic of order
$p$, then the stream function belongs to the $2p+1$ dimensional
solution space
\begin{equation}
\psi =  \sum_{\stackrel{n=1}{n\neq p}}^{+\infty} \sum_{m=-n}^n \frac{\langle \tilde{h} | Y_{nm} \rangle}{\beta_n - \beta_p}Y_{nm}(\theta, \phi) + \sum_{m=-p}^p \phi_{pm} Y_{pm}(\theta,\phi).
\end{equation}
where $\phi_{pm}$ are arbitrary coefficients. The energy and entropy
are given by
\begin{eqnarray}
E &=& E(\beta_p) + \frac{1}{2}\left( \frac{1}{R^2}-\beta_p \right) \sum_{m=-p}^p \phi_{pm}^2, \\
S &=& S(\beta_p) -\frac{\beta_p^2}{2} \sum_{m=-p}^p \phi_{pm}^2,
\end{eqnarray}
where it is assumed that in $E(\beta_p)$ and $S(\beta_p)$ the term
$n=p$ is discarded in the sum.

As in section \ref{mixedLflowsec}, the case $\lambda=\beta_1$ is only
possible when $\sqrt{\frac{4\pi}{3}} \mu = \langle h | Y_{10}
\rangle$. Then, the equilibrium flow is a superposition between the
dipolar flow of section \ref{mixedLflowsec} and the general continuum
solution described in the previous section.

\subsection{Stability of the statistical equilibria}

After some easy algebra, the second order variations of the
grand-potential functional $J[q]=S[q]-\beta E[q] - \alpha \Gamma[q] -
\mu L[q]$ are proved to be equal to
\begin{equation}
\delta^2 J = - \int_D \frac{\left( \delta q \right)^2}{2} d^2{\bf r} - \frac{\beta}{2} \int_D \left(\left({\bf \nabla} \delta \psi \right)^2 + \frac{(\delta \psi)^2}{R^2}  \right)d^2{\bf r}.
\end{equation}
As in section \ref{stabilityEGsection}, we see that if $\lambda >
\beta_1$, the equilibrium flow is a local maximum of the entropy
functional.

If there exist a couple $(n,m)$ such that $\langle h Y_{nm}\rangle
=0$, then perturbations proportional to $Y_{nm}$ destabilize the basic
flow for $\lambda < \beta_n$ while preserving the constraints at first
order. But, in general, no coefficient vanishes in the spherical
harmonic expansion and it is not sufficient to repeat as it stands the
reasoning of section \ref{stabilityEGsection} with perturbations
proportional to eigenvectors of the Laplacian since they would not
conserve the energy anymore. However, the orthogonal set of the basic
flow remains an hyperplane and one may construct many perturbations
not proportional to eigenvectors of the Laplacian. To keep the
computation of the second order variations of the free energy with
respect to the perturbation simple, we shall look for a destabilizing
perturbation in a two-dimensional subspace of this hyperplane. Let us
consider the perturbations spanned by two eigenvectors of the
Laplacian: $\delta \psi = \epsilon_1 Y_{n_1m_1} + \epsilon_2 Y_{n_2
m_2}$.  Clearly the corresponding variation of the potential vorticity
is $\delta q = \epsilon_1 \left( R^{-2} - \beta_{n_1} \right) Y_{n_1
m_1} + \epsilon_2 \left( R^{-2} - \beta_{n_2}\right) Y_{n_2 m_2}$. For
any non-zero $n_1,n_2$, this perturbation conserves the
circulation. The variation of the energy is found to be
\begin{equation}
\delta E = \epsilon_1 (R^{-2}-\beta_{n_1}) \frac{\langle h Y_{n_1m_1}\rangle}{\beta_{n_1}-\lambda}+\epsilon_2 (R^{-2}-\beta_{n_2}) \frac{\langle h Y_{n_2m_2}\rangle}{\beta_{n_2}-\lambda}.
\end{equation}
To ensure that the perturbation conserves the energy, we choose
$\epsilon_2$ such that $\delta \psi$ is orthogonal to the basic flow:
\begin{equation}\label{epsilon2eq}
\epsilon_2 = - \epsilon_1 \frac{(R^{-2}-\beta_{n_1}) (\beta_{n_2}-\lambda)\langle h Y_{n_1m_1}\rangle}{(R^{-2}-\beta_{n_2}) (\beta_{n_1}-\lambda)\langle h Y_{n_2m_2}\rangle}.
\end{equation}
It remains to see if the main flow is stable or not against this
perturbation. To that purpose we need to compute $\delta^2 J$. We find
that
\begin{equation}
\delta^2 J = -\left( \epsilon_1^2 (R^{-2}-\beta_{n_1})(\lambda-\beta_{n_1})+\epsilon_2^2 (R^{-2}-\beta_{n_2})(\lambda-\beta_{n_2})\right)/2,
\end{equation}
and with the particular choice (\ref{epsilon2eq}) for $\epsilon_2$, we
obtain the criterion:
\begin{equation}
\delta^2 J \geq 0 \iff f(\lambda) \leq 0,
\end{equation}
with 
\begin{eqnarray}
f(\lambda)=|\langle h
Y_{{n_2}{m_2}}\rangle|^2(R^{-2}-\beta_{n_2})^2(\lambda-\beta_{n_1})^3\nonumber\\
+|\langle h
Y_{{n_1}{m_1}}\rangle|^2(R^{-2}-\beta_{n_1})^2(\lambda-\beta_{n_2})^3. 
\end{eqnarray}
From
the computation of its derivative, we can see that $f$ is a
monotonically increasing function of $\lambda$. Let us define
$$\kappa^2=\frac{(R^{-2}-\beta_{n_2})}{(R^{-2}-\beta_{n_1})}\frac{|\langle
h Y_{{n_2}{m_2}}\rangle|^2}{|\langle h Y_{{n_1}{m_1}}\rangle|^2},$$
Then, $f$ vanishes for
$\lambda^*=(\beta_{n_2}+\kappa^{2/3}\beta_{n_1})/(1+\kappa^{2/3})$.
The instability condition thus reads
\begin{equation}
\delta^2 J \geq 0 \iff \lambda \leq  \lambda^*.
\end{equation}
Clearly, $\lambda^*$ lies in between $\beta_{n_1}$ and $\beta_{n_2}$,
and for $n_1=n_2=n$, $\lambda^*=\beta_n$.  In particular the choice
$n_1=n_2=1$ (with $m_1\neq m_2$) proves that all the solutions with
$\lambda < \beta_1$ are saddle points.

\subsection{The effect of the bottom topography}

We now turn to the description of the influence of an arbitrary
topography and a finite Rossby deformation radius on the structure of
the statistical equilibria. As the effect of angular momentum was
already studied thoroughly in section \ref{EGLsection}, we shall
assume in the sequel that $\mu=0$.

The interpretation of relation (\ref{psigeneral}) linking the
topography (including the Coriolis effect) and the equilibrium
streamfunction is relatively simple. The general structure of the
topography is reproduced in the equilibrium flow, with a statistical
temperature-dependent weighing of each mode. In particular, the modes
of the topography with eigenvalues close to the statistical
temperature will prevail in the final flow. In general, for a given
statistical temperature, the small-scale details of the topography
(i.e. high-order modes) will not affect the general structure of the
equilibrium flow: their amplitude in the equilibrium streamfunction
decreases relatively quickly, as $1/n^2$ where $n$ is the order of the
(degenerate) mode.

\begin{figure}
\begin{centering}
\includegraphics[width=0.5\textwidth]{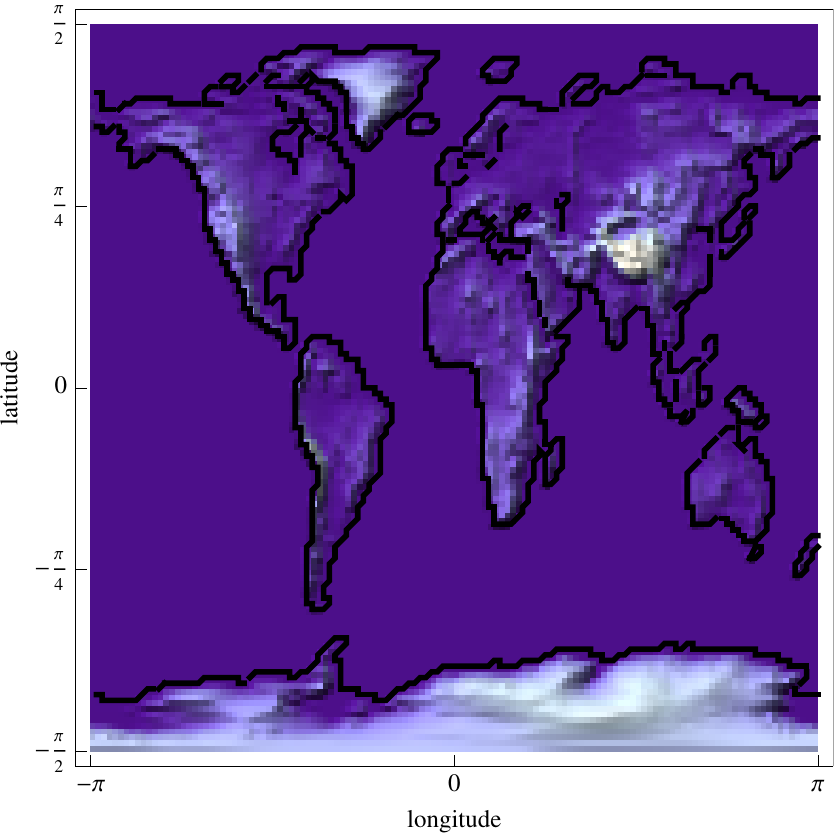}\includegraphics[width=0.5\textwidth]{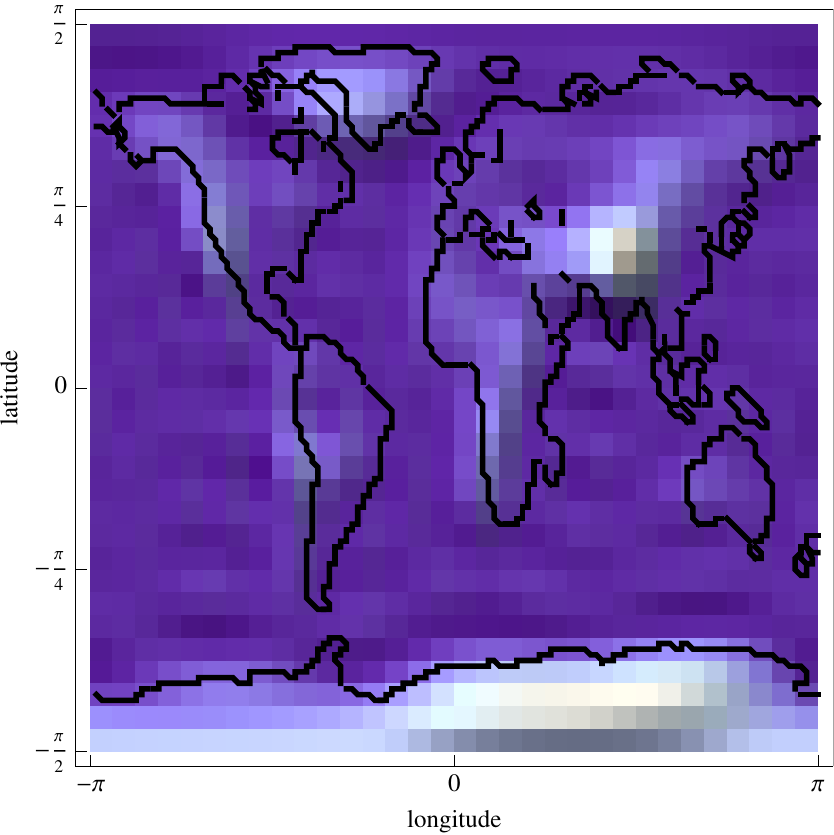}
\caption{The Earth's bottom topography (left) and its T16 spectral truncation (right).}\label{earthtopofig}
\end{centering}
\end{figure}

\begin{figure}
\begin{centering}
\includegraphics[width=\textwidth]{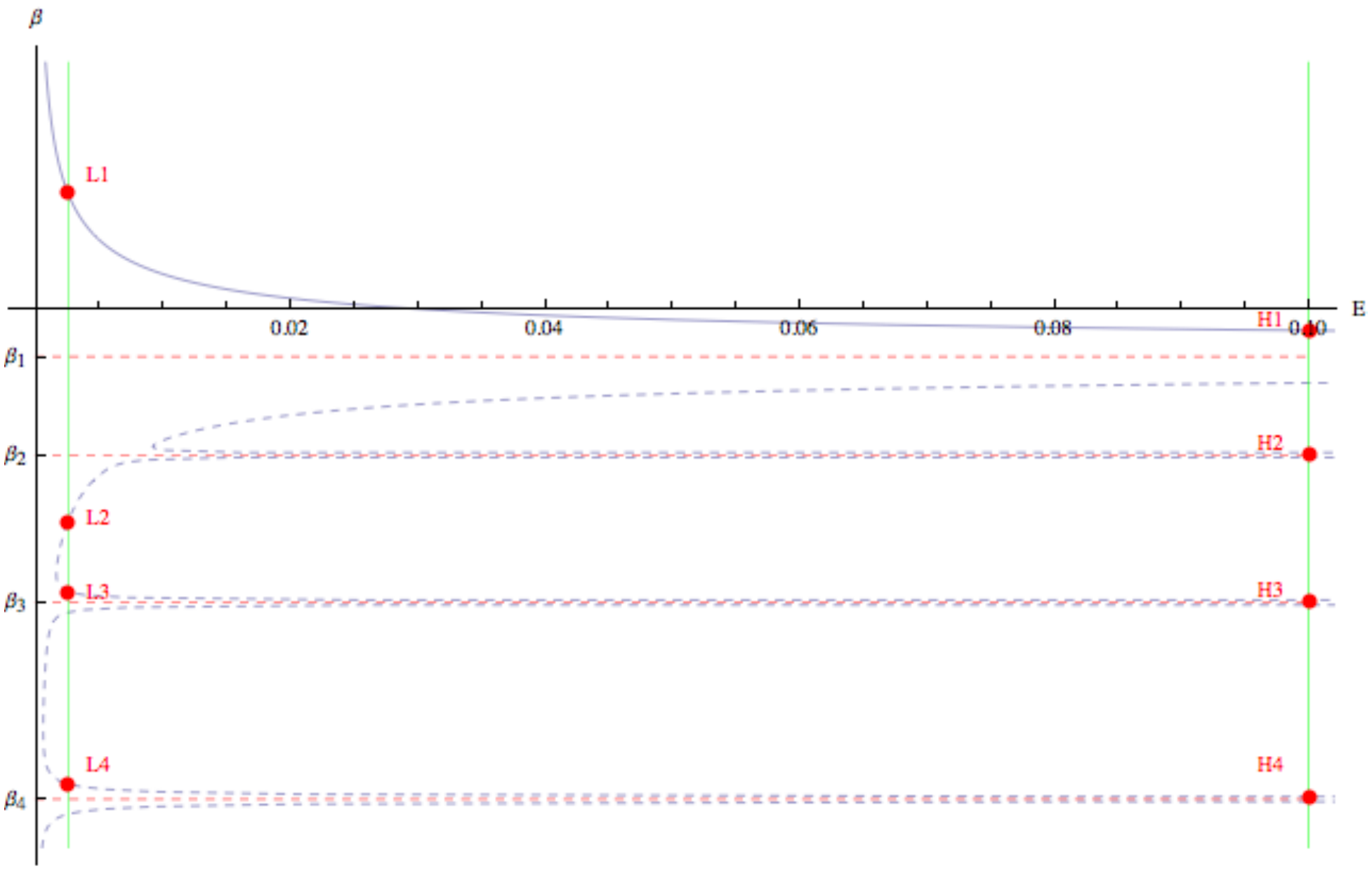}
\caption{The caloric curve $\beta(E)$ for the Earth's topography in T16 spectral truncation, with $R=\infty$. The solid blue line corresponds to entropy maxima while the solutions represented with dashed blue lines are saddle points. The position of the eigenvalues of the Laplacian is marked with red dashed lines. We show exemples of high energy flows $H1,H2,H3,H4$ in figure \ref{psitopoEbspectrum} and exemples of low-energy flows $L1,L2,L3,L4$ in figure \ref{psitopoEbcont}.}\label{caloriccurvetopoEarthfig}
\end{centering}
\end{figure}

For a more detailed discussion, we consider a specific example with
the Earth's topography in a T16 spectral truncation (see figure
\ref{earthtopofig}). Figure \ref{caloriccurvetopoEarthfig} shows the
caloric curve $\beta(E)$ for this topography in the limit of an
infinite Rossby deformation radius ($R=\infty$). This curve
essentially consists of spikes at eigenvalues of the Laplacian
superimposed on a background curve similar to the caloric curve
obtained in the absence of topography (figure
\ref{caloriccurveEGfig}). For every value of the energy, there exist
an infinity of possible statistical temperatures, but only one
corresponds to an entropy maximum (the one corresponding to $\beta>
\beta_1$). All the other solutions are saddle points of the entropy
functional, but one can imagine that they are only marginally
unstable, so that the system may remain stuck in these states for a
long time. The robustness of these saddle points has been illustrated
by Naso {\it et al.} \cite{Naso2010a} who stressed their
importance. These semi-persistent states could be relevant for climate
modeling. Indeed, the atmosphere is never in a permanent equilibrium
state. If we find large-scale structures that are, strictly speaking,
unstable but whose lifetime is, however, of the order of a few days,
such structures are fully relevant for climate modeling. They could
account for situations of atmospheric blocking. For this reason, we
shall comment on several critical points of the entropy functional
(indicated with red dots on the caloric curve of figure
\ref{caloriccurvetopoEarthfig}) in cases of high (points $H1,H2,H3,H4$) and
low energy ($L1,L2,L3,L4$) regardless of the nature of these critical
points.

\begin{figure}
\begin{centering}
\includegraphics[width=0.5\textwidth]{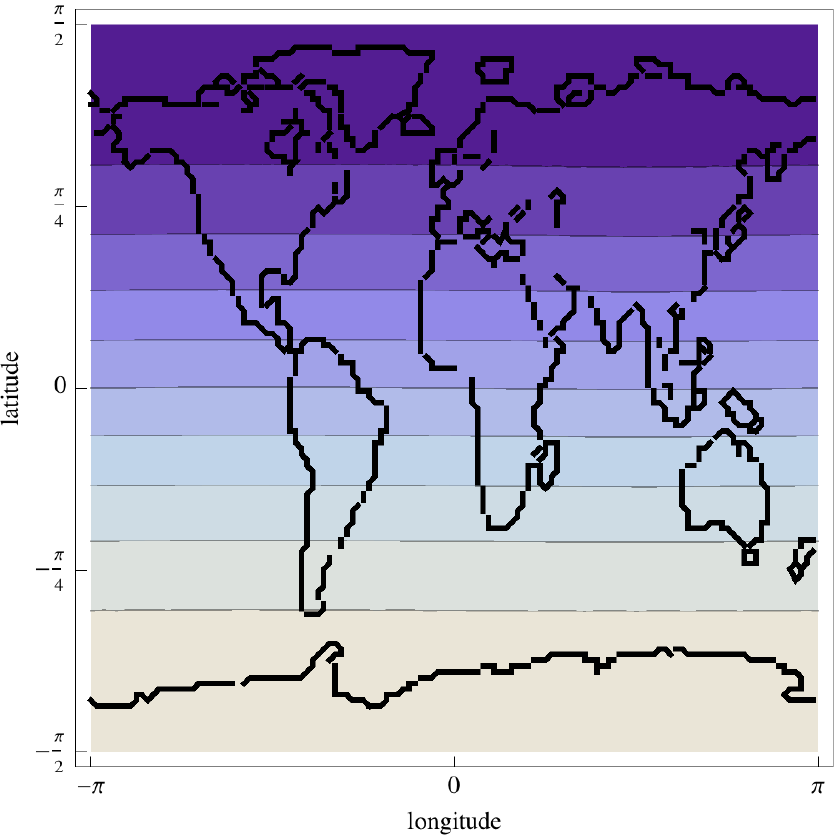}\includegraphics[width=0.5\textwidth]{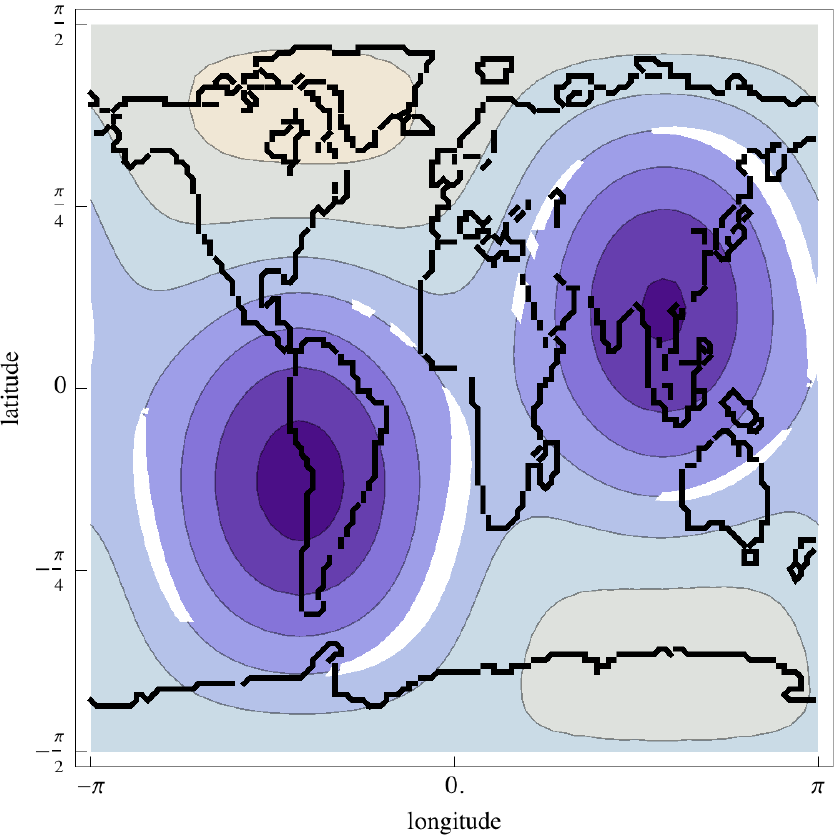}
\includegraphics[width=0.5\textwidth]{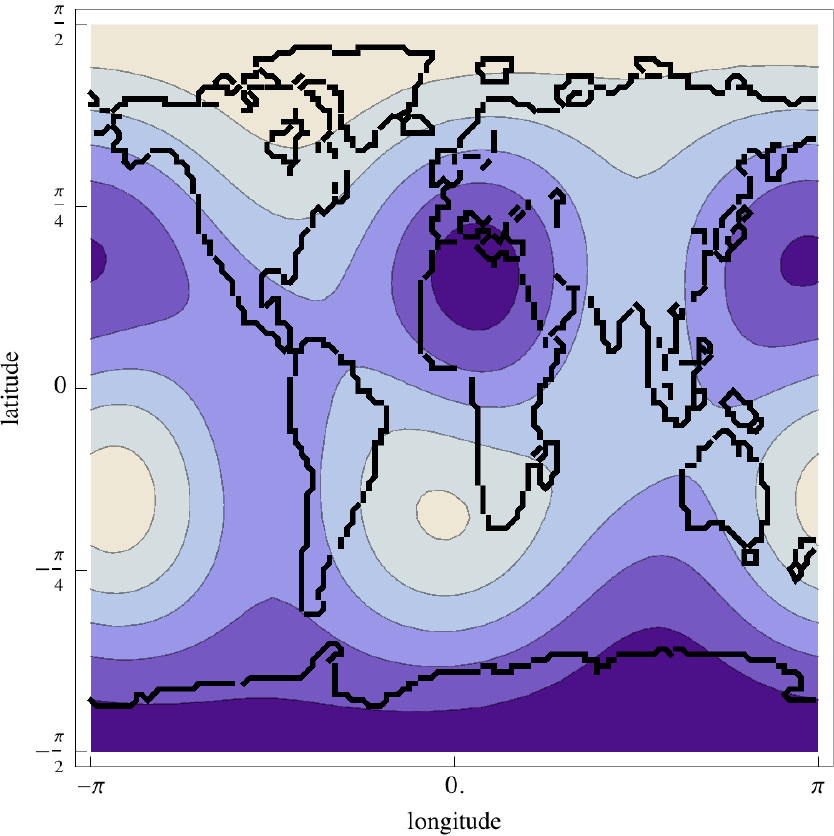}\includegraphics[width=0.5\textwidth]{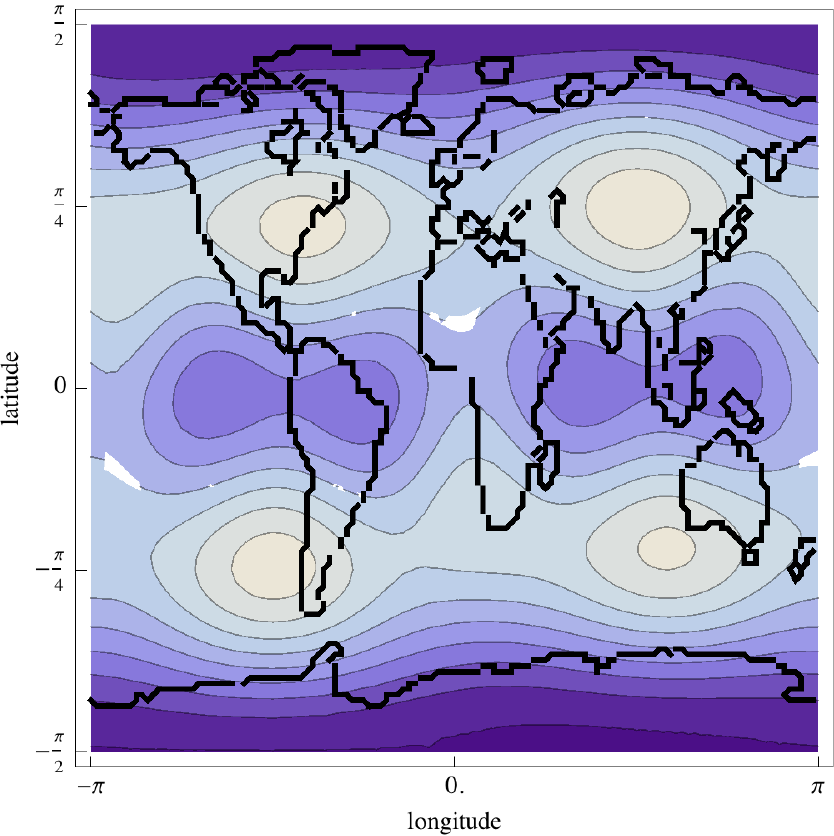}
\caption{High energy ($E=1$) flows: Equilibrium stream function for $\beta\approx\beta_1$, $\beta \approx \beta_2$, $\beta \approx \beta_3$ and $\beta \approx \beta_4$ with the Earth's topography in T16 spectral truncation, with $R=\infty$. From top to bottom and left to right, these flows are the $H1,H2,H3$ and $H4$ flows from figure \ref{caloriccurvetopoEarthfig}. }\label{psitopoEbspectrum}
\end{centering}
\end{figure}

At high energy, possible steady states for the flow are basically just
the degenerate modes. From the stability analysis of previous section,
we know that among all these modes, the only stable solution is the
one with $\beta \approx \beta_1$. Nevertheless, saddle points
corresponding to $\beta \approx \beta_n$ can be long-lived as the
system may not generate spontaneously the perturbations that
destabilize them. However, the bigger $n$ is, the more such
perturbations exist. Thus, one can expect that eigenmodes with
``large" $n$ will be less stable than eigenmodes with ``low" $n$. We
have plotted on figure \ref{psitopoEbspectrum} examples of equilibrium
streamfunctions for high-energy flows with the Earth's topography (the
first four modes are represented). For $\beta \approx \beta_1$, we
obtain the well-known solid-body rotation. The saddle solution for
$\beta \approx \beta_2$ is a quadrupole, while the states with $\beta
\approx \beta_3$ and $\beta \approx \beta_4$ (also saddles) feature
vortices belts in the middle latitudes.

\begin{figure}
\begin{centering}
\includegraphics[width=0.5\textwidth]{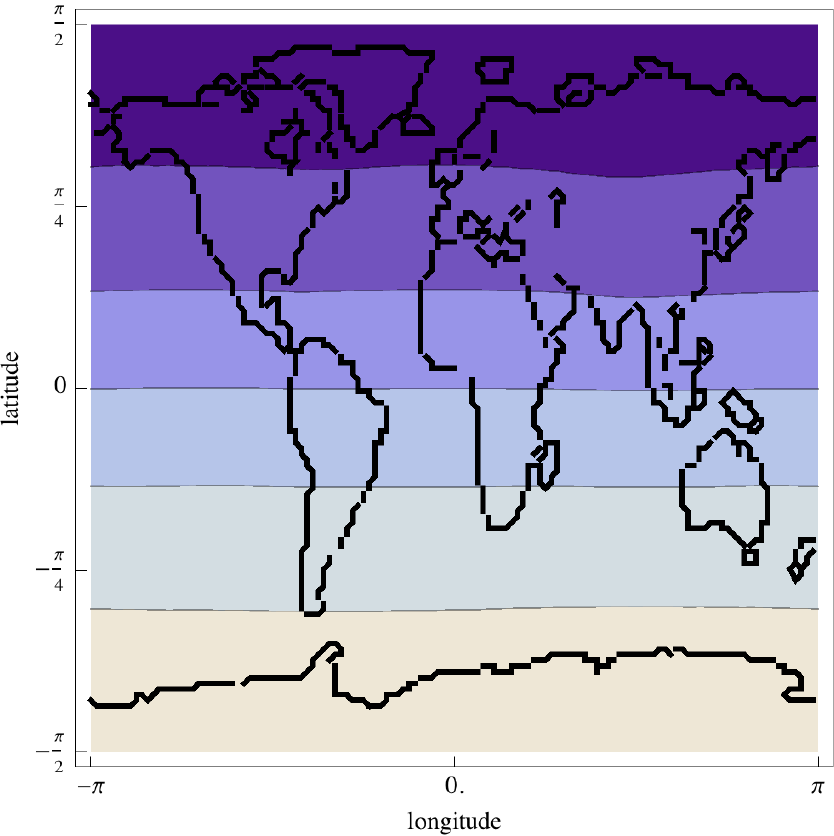}\includegraphics[width=0.5\textwidth]{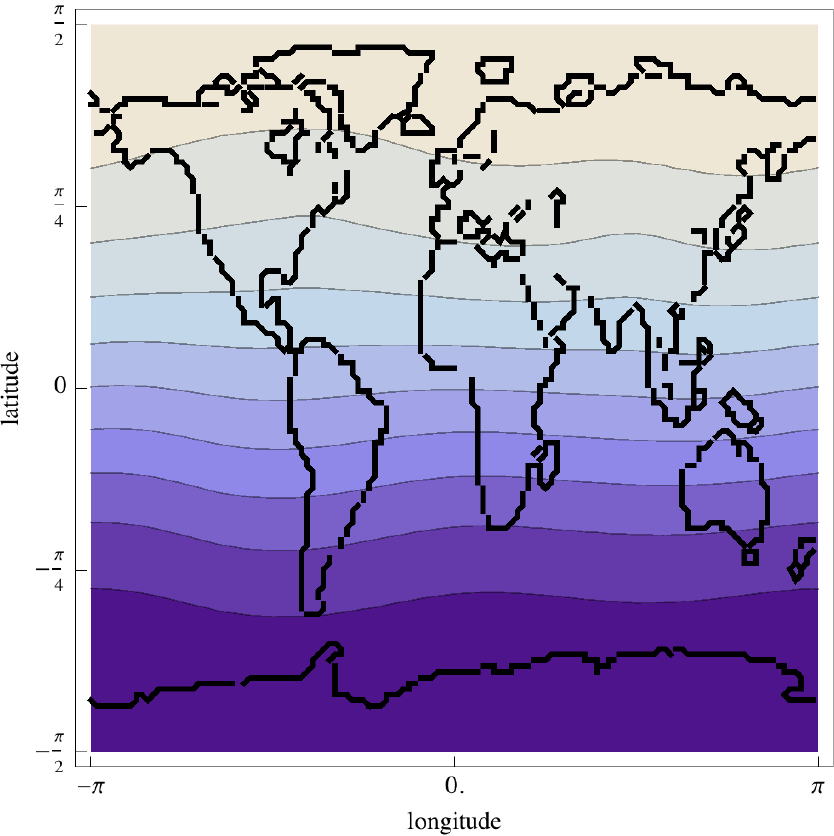}
\includegraphics[width=0.5\textwidth]{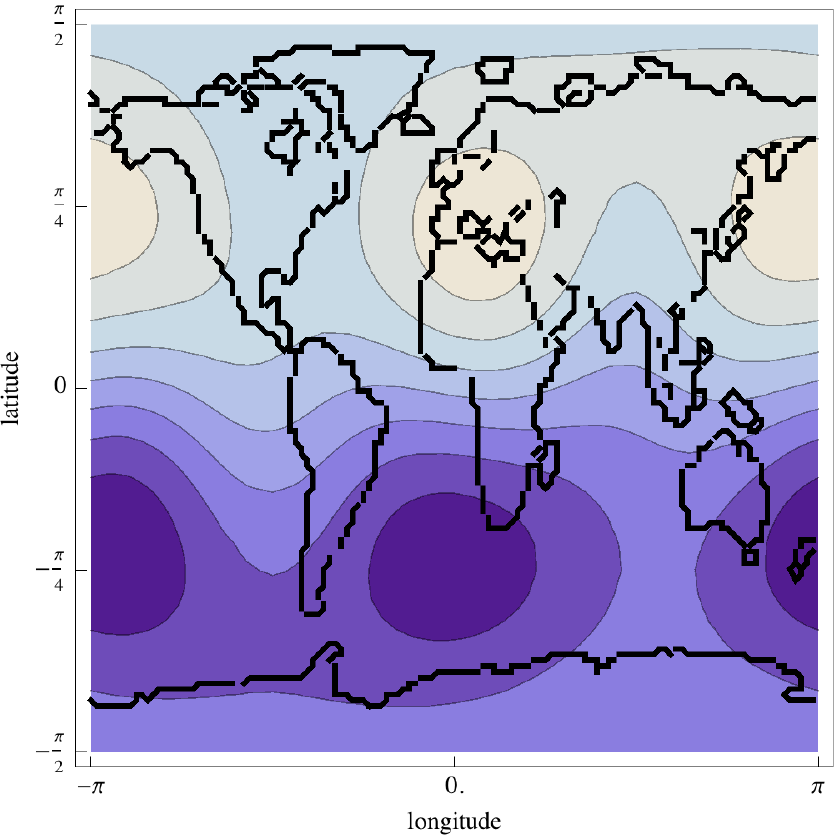}\includegraphics[width=0.5\textwidth]{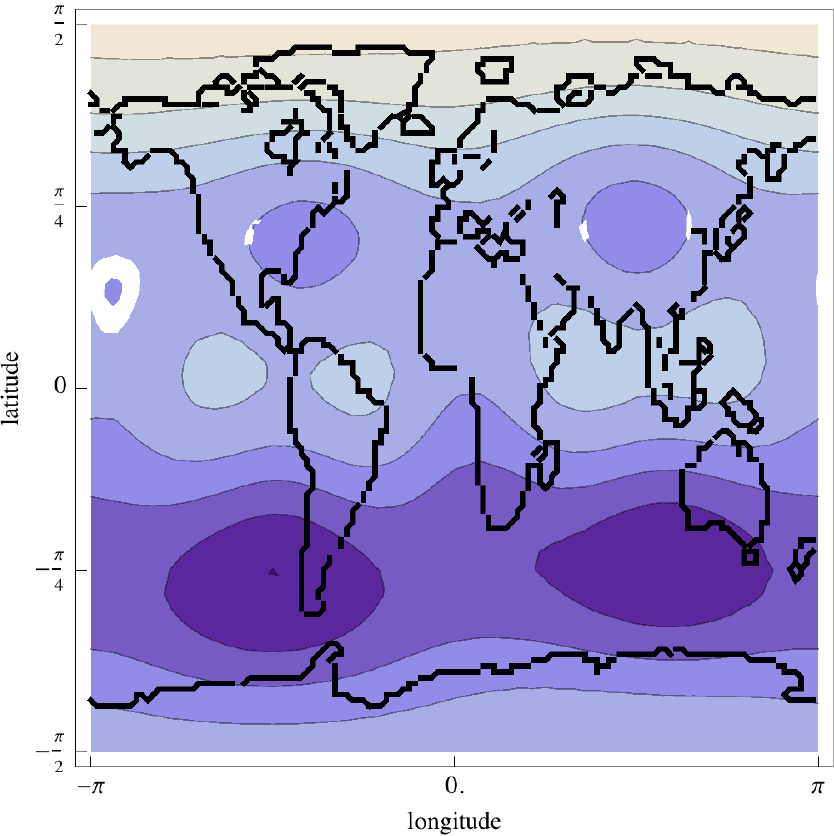}
\caption{Low energy flows ($E \approx 0.0026$): Equilibrium stream function for $\beta \approx 4.7$, 
$\beta \approx -8.7$, $\beta \approx -11.6$ and $\beta \approx -19.4$ with the Earth's topography 
in T16 spectral truncation, with $R=\infty$. From top to bottom and left to right, these flows are 
the $L1,L2,L3$ and $L4$ flows from figure \ref{caloriccurvetopoEarthfig}.}\label{psitopoEbcont}
\end{centering}
\end{figure}

At low energy, the inverse temperature $\beta$ is not constrained to
be in the vicinity of a Laplacian eigenvalue anymore. As a
consequence, the structure of the flow is less constrained and
consists of a mixing of many different modes. As examples, the
equilibrium stream functions for $\beta = 4.7, -8.7, -11.6$ and
$-19.4$ are given in figure \ref{psitopoEbcont} with the Earth's
topography. Note that, as explained before, the background solid body
rotation has a different sign depending on the sign of $\beta -
\beta_1$. Similarly to the high energy case, the solutions with $\beta
< \beta_1$ are formally unstable, but it is possible that the system
remains stuck in these states for a long time.

\begin{figure}
\begin{centering}
\includegraphics[width=0.48\textwidth]{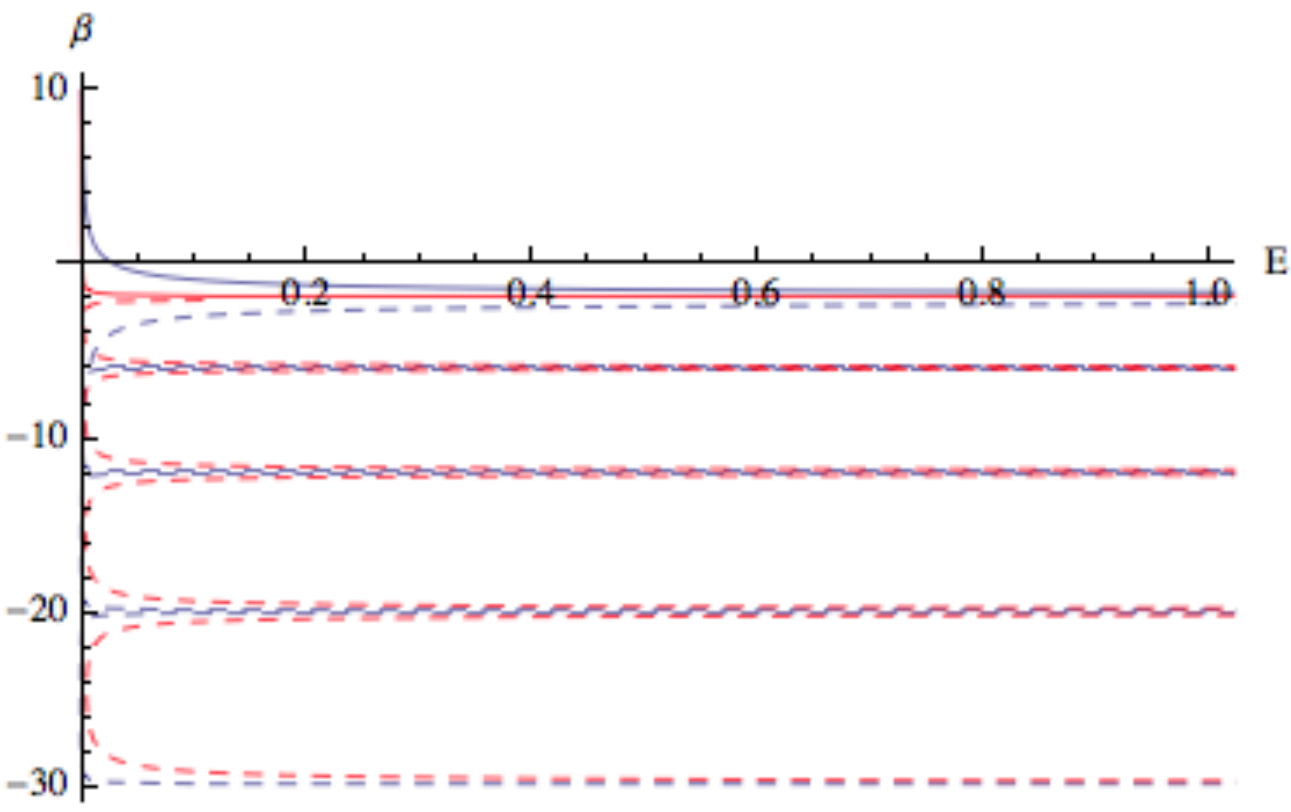}
\includegraphics[width=0.48\textwidth]{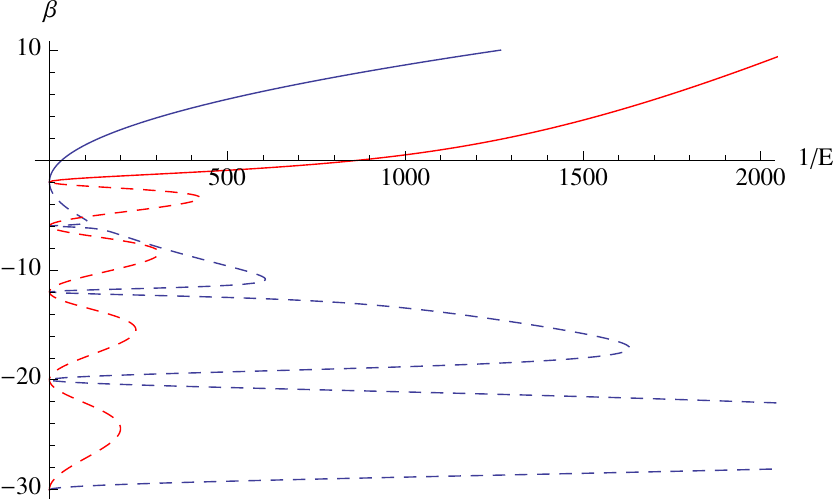}
\caption{Caloric curve $\beta(E)$ (left) and $\beta(1/E)$ (right) with $R=\infty$, 
for the Earth's topography in a T16 truncation (blue) and for a dummy topography with uniform 
spectrum and the same total power as the Earth's topography (red). The curve $\beta(E)$ (left) 
clearly shows that the influence of the topography is small at high energies as the statistical 
temperature is very close to Laplacian eigenvalues regardless of the topography. On the contrary, 
at low energy, the details of the topography are important, especially to determine which 
saddle points can be attained by the system (right).}\label{caloriccurvetopo2fig}
\end{centering}
\end{figure}

For the above discussion, we have chosen to use the topography of the
Earth to fix ideas, but in fact the qualitative features of the flow
examined here do not depend much on this particular choice. We have
also considered a dummy topography with the same truncation (T16) but
with uniform spectra, retaining the same total power as the topography
of the Earth. The resulting caloric curve is plotted on figure
\ref{caloriccurvetopo2fig} and compared to the caloric curve obtained
with the topography of the Earth. In the high energy range, 
the difference is not very important. Figure
\ref{caloriccurvetopo2fig} also shows $\beta$ as a function of $1/E$
to make apparent the differences that occur in the low energy
range. The main difference between the two curves is found in the
vicinity of $\beta_1$, as expected from the strong Coriolis domination
for the topography of the Earth. As shown in figure
\ref{caloriccurvetopo2fig} (right), the bottom topography plays an
important role at low energies, similarly to Fofonoff flows
\cite{Naso2011}. Actually, the structure of the stable equilibrium does not
depend much on the topography, even at low energy. On the contrary,
the unstable structures obtained as saddle points of the entropy, even
if they always exist, may not be reached by a flow with low energy if
the contribution of the topography is not strong enough. On figure
\ref{caloriccurvetopo2fig} (right), it is clear that the branches
between two eigenvalues of the Laplacian extend more or less towards
low energies depending on the topography. As an illustration, for the
energy of our low-energy flows L1,L1,L3,L4, the quadrupole state is
not possible.

In fact, the low-energy limit can be understood with a simple
approximation. When $\beta \to \infty$ (or equivalently, when $E\to
0$), it is possible to neglect the Laplacian (except in boundary
layers but there is none in our spherical geometry) in the mean-field
equation $\Delta \psi -\lambda \psi = \alpha +h +\mu \cos \theta$. We
obtain $\lambda \psi= -\alpha-h-\mu\cos\theta$, which means that the
flow is a superposition of a solid-body rotation term $-(\mu/\lambda)
\cos \theta $ due to the conservation of the angular momentum, and a
term $-h/\lambda$ directly fixed by the topography. Again, this is
similar to the case of Fofonoff flows studied in \cite{Naso2011}.

\subsection{The role of the Rossby deformation radius}

\begin{figure}
\begin{centering}
\includegraphics[width=0.7\textwidth]{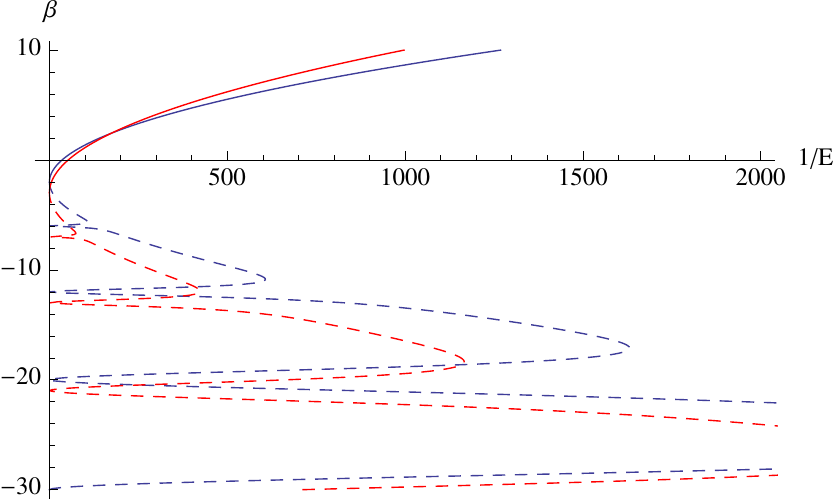}
\caption{Caloric curve $\beta(1/E)$ for the Earth's topography in a T16 spectral truncation, 
with infinite Rossby deformation radius (blue) and with $R=1$ (red). The effect of the finite 
Rossby deformation radius is mainly seen for low values of the energy.}\label{caloriccurveRossbyfig}
\end{centering}
\end{figure}

In the previous section, we have analyzed the effect of arbitrary
topographies on the equilibrium states, but only in the limit of
infinite Rossby deformation radius. In fact, the mean field equation
for a finite Rossby deformation radius is the same as that for
infinite $R$ if we make the change of variable for the statistical
temperature $\lambda = \beta+1/R^2$. As a consequence, the set of
equilibrium states in itself is unchanged. Indeed, as equation
(\ref{psigeneral}) shows, the set of stream function spanned as we
vary $\lambda$ is the same as that obtained by varying $\beta$ in the
limit $R=+\infty$.  On the contrary, the expressions for the energy
(equation (\ref{energygeneraleqn})) and entropy (equation
(\ref{entropygeneraleqn})) depend directly on $R$ and not only on
$\lambda$. Figure \ref{caloriccurveRossbyfig} compares the caloric
curve obtained for $R=\infty$ and for $R=1$ (with $\langle \psi
\rangle =0$ or, equivalently, $\Gamma=\langle h \rangle$). The main
effect of the finite Rossby deformation radius is to shift the curve
towards negative statistical temperature values, as expected from the
identity $\lambda = \beta+1/R^2$. The qualitative picture discussed
above is not modified by the Rossby deformation radius if we replace
everywhere $\beta$ by $\lambda$. Note that for a fixed statistical
temperature $\beta$, the energy is generally higher for a finite
Rossby deformation radius.

\section{Discussion}\label{discussionsection}

Let us summarize the results obtained in the previous sections.
The statistical equilibrium states (in the sense as those corresponding to a linear
$q-\psi$ relationship) of the quasi-geostrophic equations with
conservation of energy and circulation on a rotating sphere are as
follows:
\begin{itemize}
\item For a smooth (without bottom topography) sphere without taking 
into account the conservation of the vertical component of the angular
momentum: the only stable state is a counter-rotating solid-body
rotation. There is always an unstable equilibrium state which is a
co-rotating solid-body rotation. Depending on the energy, one may also
have unstable degenerate eigenmodes (since the coefficients are
undetermined, these can be any combination of the corresponding
spherical harmonics), which are saddle points of the entropy.

\item For a smooth sphere taking into account conservation of the 
vertical component of the angular momentum, there are two
possibilities: if the energy and angular momentum values satisfy the
relation $E=E^*(L)$, we have a solid-body rotation with angular
velocity $\Omega_*=3L/2$. In particular, both counter-rotating and co-rotating
rotations are possible as stable equilibria. Otherwise, the
statistical equilibrium is a dipole: in this case, we observe a
spontaneous symmetry breaking associated with a second order phase
transition. Again there are possibilities of unstable saddle points
states as in the previous case.

\item For a sphere with a bottom topography, the equilibrium states 
are unchanged, but there are extra saddle points states featuring
higher-order spherical harmonics ($n\geq 2$) that are more constrained
(due to the topography) than when they appear as degenerate solutions
of the Helmholtz mean-field equation. At high
energies, the topography has a negligible effect, while it becomes
important at low energies.

\end{itemize}

One of the original motivations of the theories of statistical
mechanics for two-dimensional turbulent flows \cite{Chavanis1996,
Chavanis1998b} was to account for the observed tendency for the flow
to organize into coherent structures, such as monopoles, dipoles,
tripoles, quadrupoles, etc
(\cite{McWilliams1984,McWilliams1990a,McWilliams1990b,Brachet1986,
Brachet1988,Benzi1986,Benzi1988,Babiano1987,Santangelo1989,Schneider2005}).
The numerical simulations suggesting these large-scale structures, as
well as the theoretical attempts to obtain them as statistical
equilibria, focus on planar flow cases with either fixed or periodic
boundary conditions, and flows in a $\beta$-channel
\cite{Rhines1975}. In the present study, we find that the combined
effect of rotation and spherical geometry partially annihilates these
coherent structures: in the absence of a bottom topography, the
equilibrium flow is purely zonal. This prediction from the theory is
confirmed by early results obtained in numerical simulations of 2D
turbulent flows on a sphere \cite{Cho1996,
Yoden1993,Williams1978,Basdevant1981}. However, if we impose an
additional constraint of angular momentum conservation, we recover a
dipolar equilibrium state (for some values of the external
parameters). Interestingly, in this case the statistical mechanics
presents a spontaneous symmetry-breaking feature: we obtain a set of
equilibrium states with a phase factor as a free parameter. For each
particular equilibrium state, the axial symmetry is broken but the
action of rotations around the vertical axis leaves the set of
equilibrium states globally invariant. This spontaneous symmetry
breaking appears due to the presence of a second order phase
transition, which occurs in both the microcanonical and the grand-canonical ensembles. 
It is also noteworthy that the thermodynamical properties
of the system are quite unusual \cite{Herbert2011c}. The
microcanonical and grand-canonical ensembles are equivalent, but only
marginally so: the entropy functional is a plane and, as such, it is
concave but it is also convex. The statistical temperature does not
depend on the energy in the dipole phase, and can be anything greater
than $\beta_1$ in the solid-body rotation phase. Besides, the specific
heats $\partial^2 {\cal S}/\partial E^2$ and $\partial^2 {\cal
S}/\partial L^2$ both vanish.

In direct simulations on a non-rotating sphere, conservation of all
the components of angular momentum leads to the formation of coherent
structures, while in the presence of rotation, when only one component
of the angular momentum is conserved, zonal structures emerge. In
fact, carrying out the statistical mechanics procedure developed in
this study with all the components of the angular momentum conserved
(and thus vanishing rotation) indeed yields coherent structures
similar to those observed in numerical simulations. The two extra
conserved quantities add terms proportional to $\sin \theta \cos \phi$
and $\sin \theta \sin \phi$, which lead to a flow similar to the
dipole obtained in section \ref{mixedLflowsec}, except that the phase
is now fixed by the angular momentum constraint. Besides, this flow
perdures for external parameters varying in a wide range, whereas it
only occurs for $\beta=\beta_1$ and $\mu=-2\Omega$ in section
\ref{mixedLflowsec}.

On the rotating sphere, as summarized above, the stable equilibrium
flow does not present complex vortices, except in one case if we take
into account conservation of angular momentum. But as shown on figure
\ref{psitopoEbspectrum} in the presence of a bottom topography, many
coherent structures can be realized as unstable saddle points states
with a statistical temperature close to a Laplacian eigenvalue. Those
corresponding to low-order eigenvalues (like the quadrupole) are
likely to be only weakly unstable, since only low-order perturbations
can destabilize them. They could account for atmospheric blocking
where a vortex persists for a few days and is finally destabilized and
disappears.

Finally, also note that even in the absence of a bottom topography and
without conservation of angular momentum, coherent vortices can be
obtained as degenerate modes ($\beta \in \mbox{ Sp } \Delta$). The
resulting stream function then resembles the examples in figure
\ref{psitopoEbspectrum}, except that the coefficients are not
determined in any way. Thus, virtually any combination of the
eigenvectors for this eigenvalue is an acceptable equilibrium stream
function. However, as pointed out before, these states are not stable,
even though they can be long-lived, especially low-order modes.

It would be of great interest to compute the statistical equilibrium
obtained from the theory for realistic values of the constraints
(kinetic energy, angular momentum) and to compare it to observations
and predictions from dynamical models. In the case of the Earth, this
was done by Verkley and Lynch \cite{Verkley2009b} in the framework of
the Kraichan energy-enstrophy theory for a spectrally truncated model,
with partial agreement. We shall report elsewhere the results obtained
by doing so with the theory developed here. Note that
\cite{Verkley2009b} considers some simple representation of forcing
and dissipation. Although the results obtained are very encouraging,
it is difficult to justify rigorously the inclusion of forcing and
dissipation, as the Liouville theorem does not automatically hold in
this case. However, in real flows, forcing and dissipation do not
equilibrate locally, and they play an important role in the theory of
geophysical fluids. Nevertheless, there are some known cases where
quasi-stationnary states reached by a forced-dissipated system are
well approximated by some equilibrium states of a system with no
forcing and no dissipation. An important example is the von Karman
flow \cite{Naso2010b,Monchaux2006}. This may indicate that, at least
in some cases, the non-equilibrium attractors may remain in the
vicinity of some equilibrium states. However, the general question of
the relevance of equilibrium approaches to non-equilibrium problems is
far from being understood.


\section{Conclusion}\label{conclusionsection}

In this paper, we have applied the Miller-Robert-Sommeria statistical theory
for perfect inviscid fluids to the general circulation on a rotating
sphere. The large-scale circulation is modelled by a one-layer
(barotropic) quasi-geostrophic flow and the potential vorticity plays
the role of the vorticity in the MRS theory. If we
only consider the conservation of the fine-grained enstrophy among the
infinite class of Casimirs, the maximization of the MRS entropy at
fixed energy, circulation, angular momentum and fine-grained enstrophy is equivalent,
for what concerns the mean field, to the minimization of the
coarse-grained enstrophy at fixed energy, circulation and angular momentum
\cite{Naso2010a}. This leads to a linear $q-\psi$ relationship. 
Furthermore, the fluctuations around the mean field state are
gaussian. We have shown that the mean field equation can be
solved analytically in a very simple way due to the geometry of the
domain and the fact that the Coriolis parameter is an eigenvector of
the Laplacian on the sphere. Ignoring the conservation of angular
momentum, we have found that the stable statistical equilibrium flow is a
counter-rotating solid-body rotation. Taking the conservation of angular momentum into account, we have obtained 
two qualitatively different
equilibrium flows: solid-body rotation (reminiscent of the previous
case) or dipole. In the latter case, the axial symmetry is
spontaneously broken and the system displays a second order phase
transition. Finally, we have shown that the equilibrium quasi-geostrophic
flow on a sphere with arbitrary bottom topography has the same
background structure with topography-induced modes superimposed. In
particular, unstable saddle points modes with multipole vortex
structures appear, even though they were already possible as
degenerate modes in the absence of a bottom topography.  Albeit
obtained with a relatively simple model, these results suggest that
statistical mechanics constitutes a nice and efficient framework for
theoretical studies of large-scale planetary circulation.  They also
provide a strong incentive to generalize the statistical mechanical
methods to more realistic models of the atmosphere and oceans.


\appendix

\section{Solid-body rotations}\label{solidbodyrotationsappendix}

There is a specific type of flow which is of central importance in
this study: solid-body rotations. These correspond to the case when
all the fluid revolves around the axis of rotation of the sphere,
exactly as if it were a solid. Under these conditions, the stream
function reads
\begin{equation}
\psi =\Omega_* \cos \theta,
\end{equation}
where $\Omega_*$ is the angular
velocity of the solid-body rotation. Clearly for this specific flow,
all the dynamical invariants are not independent, as all the dynamical
quantities depend only on $\Omega_*$. In fact, straightforward
computations lead to
\begin{eqnarray}
E&=&\frac{{\Omega_*}^2}{3},\\
L&=&\frac{2}{3}\Omega_*,\\
\Gamma_2&=&\frac{1}{3} (\Omega+\Omega_*)^2.
\end{eqnarray}
We thus introduce the function $E^*(L)=3L^2/4$ which gives the energy of a
solid-body rotation with angular momentum $L$. In \ref{ELappendix}, we
show that the energy of any flow on the sphere is always greater than
the energy of a solid-body rotation with the same angular momentum.
Alternatively, we can define the functions $L_+^*(E)=\sqrt{{4E}/{3}}$
and $L_-^*(E)=-\sqrt{{4E}/{3}}$. The angular momentum of a solid-body
rotation in the positive direction with energy $E$ is $L_+^*(E)$,
while $L_-^*(E)$ is the angular momentum of a solid-body rotation in
the negative direction with energy $E$. Clearly, specifying the energy
of the solid-body rotation is not enough, one needs to know in
addition the direction of rotation, hence the two functions $L_+^*$
and $L_-^*$.

Note that the relation between the energy, the angular momentum, and
the enstrophy which arises as the thermodynamic equilibrium entropy in
the text is already fixed by the dynamical constraints in the case of
a solid-body rotation: $-\Gamma_2/2=-2\Omega^2/3-2E-2\Omega L$.

\section{Minimum energy for a flow with given angular momentum}\label{ELappendix}

For quasi-geostrophic flows on a rotating sphere, the kinetic energy reads 
$E=\langle u^2+v^2 \rangle/2$ where $u$ and $v$ are the components of the velocity. 
The vertical component of the angular momentum is $L=\langle u \sin \theta \rangle = \langle (q-f) \cos \theta \rangle$. The Cauchy-Schwarz inequality immediately yields 
\begin{equation}
\left( \int_{S^2} u \sin \theta dS \right)^2 \leq \frac{8\pi}{3} \left( \int_{S^2} u^2 dS \right),
\end{equation}
and consequently,
\begin{equation}
E \geq \frac{3}{4} L^2.
\end{equation}
Therefore, we always have $E\geq E^*(L)$. The lower bound for the energy is reached 
for a solid-body rotation (see \ref{solidbodyrotationsappendix}). We can also say 
that, for a given energy $E$, the angular momentum $L$ must satisfy $L_-^*(E) \leq L \leq L_+^*(E)$.

Another derivation of this inequality sheds more light on its physical interpretation. 
For a given value of the angular momentum $L$, it is always possible to find a 
reference frame in which $L'=0$. Indeed, in a reference frame ${\cal R}'$ rotating 
with angular velocity $\Omega'$ relative to the Earth's rotation, straightforward 
computations give
\begin{eqnarray}
L'(\Omega')&=&L-\frac{2}{3} \Omega',\\
E'(\Omega')&=&E-L \Omega' +\frac{1}{3} \Omega'^2,
\end{eqnarray}
where $E'$ is the energy in ${\cal R}'$. Clearly, the value of $\Omega'$ such that 
$L'=0$ is $\Omega'=\frac{3}{2}L$, and it is also the value for which $E'$ 
is a minimum. Thus
\begin{equation}
E'\left (\frac{3}{2 L}\right)=E-\frac{3}{4}L^2 \geq 0,
\end{equation}
so that we again obtain $E\geq E^*(L)$.

Finally, the shortest way to this inequality is perhaps to decompose the fields on spherical harmonics so that
\begin{equation}
\omega=\sum_{n=0}^{+\infty}\sum_{m=-n}^n \omega_{nm} Y_{nm},\qquad 
 \psi= \sum_{n=0}^{+\infty}\sum_{m=-n}^n \frac{\omega_{nm}}{\beta_n} Y_{nm}, 
\end{equation}
and
\begin{equation}
E=-\frac{1}{8\pi} \sum_{n=0}^{+\infty}\sum_{m=-n}^n \frac{| \omega_{nm} |^2}{\beta_n}.
\end{equation}
Clearly, $L= \langle \omega | Y_{10} \rangle /\sqrt{12\pi} = \omega_{10} / \sqrt{12\pi}$ 
and $E\geq |\omega_{10}|^2/(-8\pi\beta_1)$ so that, again, $E\geq E^*(L)$. This proof makes 
it evident that the inequality only means that the energy is always at 
least the energy contained in the solid-body rotation mode, which is directly fixed by the angular momentum.

For fixed enstrophy, the classical Fjortoft argument \cite{Fjortoft1953} also gives an upper bound on the energy:
\begin{equation}\Gamma_2=4\Omega L + \frac{4}{3}\Omega^2 +\frac{1}{4\pi}\sum_{n,m} | \omega_{nm} |^2,
\end{equation}
so that $E \leq \sum_{n,m} | \omega_{nm}|^2 /(-8\pi \beta_1) = (\Gamma_2-4\Omega L -4\Omega^2/3)/4$. Finally,
\begin{equation}
 \frac{3}{4} L^2 \leq E \leq \frac{\Gamma_2}{4}-\Omega L -\frac{\Omega^2}{3},
 \end{equation}
where all the inequalities become equalities in the case of a solid-body rotation.

\clearpage

\section*{References}

\end{document}